\documentclass[iop,appendixfloats]{emulateapj}

\newcommand{\figurepath}{.}

\newcommand{\cs}{c_{\rm s}}

\newbox\grsign \setbox\grsign=\hbox{$>$}
\newdimen\grdimen \grdimen=\ht\grsign
\newbox\laxbox \newbox\gaxbox
\setbox\gaxbox=\hbox{\raise.5ex\hbox{$>$}\llap
     {\lower.5ex\hbox{$\sim$}}}\ht1=\grdimen\dp1=0pt
\setbox\laxbox=\hbox{\raise.5ex\hbox{$<$}\llap
     {\lower.5ex\hbox{$\sim$}}}\ht2=\grdimen\dp2=0pt

\newcommand{\sn}{\bf \color{black} }

\shorttitle{Turbulence \& Stellar jets}
\shortauthors{Sheikhnezami \& Abgharian}
\usepackage[usenames,dvipsnames]{color}
\usepackage{amsmath}
\usepackage{float}
\usepackage{ctable} 
\usepackage{color}
\begin{document}

\title{Turbulence driven by stellar jets, the possibility and the efficiency
}
\author{Somayeh Sheikhnezami
\altaffilmark{1} 
\&
        Mohsen Abgharian-Afoushteh
 \altaffilmark{1},
}
\altaffiltext{1}{School of Astronomy, Institute for Research in Fundamental Sciences (IPM), P.O. Box 1956836613, Tehran, Iran}
\email{snezami@ipm.ir \& m.abghari@outlook.com}                                   

\begin{abstract}
We investigate the feedback of the stellar jets on the surrounding interstellar gas based on 2D and 3D
simulations applying HD and MHD module of PLUTO 4.2 code.
The main question we address is whether the stellar jet can be considered as a turbulence driver 
into the interstellar gas.
In addition, we investigate the most effective circumstances in which the driven turbulence is larger and can
survive for a longer time scale in the ambient gas.
 We present a case study of different parameters runs including the jet Mach number, the initial jet velocity field 
and the background magnetic field geometries and the interacting jets.
Also, we study the environmental effects on the jet-gas interaction by considering
the non-homogeneous surrounding gas containing the clumps in the model setup.
 Among different setups, we find that for (1) a higher jet Mach number, (2) a rotating jet ,(3) a jet propagating in a magnetized environment,
(4)a jet propagating in a non-homogeneous environment,  and (5) the interacting jets
more fluctuations and random motions are produced in the entrained gas which 
can survive for a longer time scale.
In addition, we perform the 3D simulations of jet-ambient gas interaction 
and we find that the amount of (subsonic-supersonic) fluctuations
increases compared to the axisymmetric run
and the entrained gas gains higher velocities in a 3D run.
In total, we confirm the previous finding that the stellar jets can transfer the 
turbulence on neighboring regions and are not sufficient 
drivers of the large-scale supersonic turbulence in molecular clouds.

\end{abstract}
\section{Introduction}
Astrophysical jets are produced by many objects over a wide range of luminosity and spatial scale, 
from young stellar objects to micro-quasars, and active galactic nuclei (AGN).
Jets and outflows carry the mass, energy and angular momentum and have major impacts on
the dynamical evolution of their host systems.
Jets and outflows besides supernovae, stellar winds, and spiral arms
are known as sources providing energy input and turbulence in
molecular clouds and star-forming regions \citep{2004RvMP...76..125M}.
Supersonic turbulence is an essential ingredient which can put the constraints on the star formation process.
However, the possible dual effects of the turbulence on star formation is controversial.
It can compress the material and therefore change the Jeans parameter and provide the condition to form stars.
On the other hand, the supersonic turbulence is able to disrupt the clump and decreases
the star formation \citep{2007ARA&A..45..565M}.

Observational studies show that outflows can transfer sufficient energy to account for cloud turbulent
energy \citep{1996ApJ...473..921B, 2000A&A...361..671K, 2003AJ....126..893B, 2005ApJ...632..941Q}. 
One of the essential aspects of studying jet feedback on the host systems is the corresponding scale of the system.
The jet feedback can be studied on the scale of star-forming clusters or the entire molecular 
cloud ($l > 10^4 AU$) which is not the concerns of the current paper.
In addition, the jet feedback can be constrained to the jet launching area ($l < 10 AU$) including the launching 
process of jet besides the impact on the host system.
There are numbers of MHD simulations have investigated the jet launching process 
\citep{2002ApJ...581..988C, 2007A&A...469..811Z, 2010A&A...512A..54C, 2012ApJ...757...65S,2014ApJ...796...29S, 2016ApJ...825...14S, 
2015ApJ...814..113S,2018ApJ...861...11S}. However, the jet launching process in combination with the jet feedback needs to be studied deeply. 
Moreover, the jet feedback can be constrained to the larger scale ($l < 10^4 AU$) in which the impact of jets on the  
infalling envelopes is considered \citep{2000ApJ...530..923D}. 

The first study of the stellar outflow feedback was performed by \cite{2000ESASP.445..457M}.
They discuss that the stellar outflow can drive turbulent motions.
Later, \citet{2007ApJ...668.1028B} performed numerical simulations of stellar jets and discuss that the collimated jets from young stellar
objects are unlikely drivers of large-scale supersonic turbulence in the molecular cloud.
In addition, \cite{2009ApJ...692..816C} have carried out simulations of a single outflow interacting with a turbulent medium.
They conclude that outflow-driven cavities are able to re-energize turbulent motions in their immediate environment
provided that such turbulent motions already exist to disrupt the cavity.
There is a vast of literature investigating the driving of the turbulence in their surrounding area at different scales
\citep{2009ApJ...695.1376C, 2009ApJ...704..137J, 2010ApJ...709...27W, 2010ApJ...722..145C, 2012ApJ...747...22H, 2014ApJ...790..128F}.
However, the efficiency of driving the turbulence by the stellar jet is not completely known.

In the current paper we aim to perform the case study of different stellar jet parameters to investigate 
the most efficient circumstances of driving the turbulence to the surrounding gas.
Here we are mainly interested in high-velocity jets as a potential source of supersonic turbulence and we 
do not consider other physical processes like radiation or heating and cooling of the system.

Compared to the previous works, we perform a wider case study both in HD and MHD regimes and we discuss the role of various physical parameters 
like velocity components or the background magnetite field geometry which have not been studied in details in previous works.
We discuss the turbulence generally and we are not constrained just to the supersonic turbulence.
The main question we address is whether the stellar jet can deliver the turbulence into the interstellar medium.
In addition, we aim to investigate the most appropriate circumstances in which the larger fraction of turbulence can be transferred to the 
surrounding gas and survive for a longer time scale.

The paper is organized as follows; 
in section 2 we describe the model setup, initial and boundary conditions.
in section 3 we present and discuss the different parameters runs.
and in section 4 we summarize the results.

\section{Model setup}
In this paper, we focus on the interaction of the jet with the interstellar gas to investigate
the efficiency of driving the (subsonic-supersonic) turbulence by the stellar jet into the ambient gas.
In particular, we study the global effects of the propagation of a formed jet far from the source (star)
and thus not constrained by the gravity of the source.
\subsection{Initial conditions}
For our numerical simulations, we use PLUTO code (Astrophysical gas dynamics code, version 4.2, \citet{2007ApJS..170..228M, 2012ApJS..198....7M}) 
solving the time-dependent, HD and ideal MHD equations, namely for the conservation of mass, momentum, and energy:

\begin{equation}
\frac{\partial\rho}{\partial t} + \nabla \cdot(\rho \vec v)=0,
\end{equation}
\begin{equation}
\frac{\partial(\rho \vec v)}{\partial t} + 
\nabla \cdot \left(\vec v \rho \vec v - \frac{\vec B \vec B}{4\pi} \right) + \nabla \left( P + \frac{B^2}{8\pi} \right)
+ \rho \nabla \Phi = 0,
\end{equation}

Here, $\rho$ is the mass density, $\vec v$ is the velocity, $P$ is 
the thermal gas pressure, $\vec B$ stands for the magnetic field, 
and $\Phi$ denotes the gravitational potential which is ignored in 
our model setup since we study the jet area far from the central object.

The total energy density is
\begin{equation}
e = \frac{P}{\gamma - 1} + \frac{\rho v^2}{2} + \frac{B^2}{2} + \rho \Phi.
\end{equation}
The gas pressure follows a polytropic equation of state $P = (\gamma - 1) u$ with 
$\gamma = 5/3$ and the internal energy density $u$.
Our model setup includes the jet environment and the surrounding gas in which
the jet area has a lower density but is in pressure equilibrium with the ambient gas.
{The evolution of the magnetic field (for MHD setup)is described by the induction equation (in absence of the diffusive term),
\begin{equation}
\frac{\partial \vec B}{\partial t} - \nabla\times (\vec v \times \vec B ) = 0.
\end{equation}

Including the heating or cooling processes is beyond the goals of the current paper, and 
we will study the jet propagation in the non-ideal MHD regime, in a future paper.


\begin{table*}
	\centering
	\caption{Characteristic parameters of our simulation runs. 
    Here  $\rho_j$ is the jet density, $v_{r}$ is the radial velocity of jet, $v_{\phi}$ is the jet rotation ,
    $\rho_a$ is the ambient gas density, $\delta = \frac{\rho_j}{\rho_a} $ is  the ratio of the jet density to the
    ambient gas density, $ M = \frac{V_{\rm j}}{c_{\rm s}}$is the Mach number
   which is the ratio of the jet velocity to the sound speed, $B_{\rm r}$ is the radial, $B_{\rm z}$ is the vertical and $B_{\rm phi}$ is the toroidal
   background magnetic field.
    }
	\label{tab: 1st}
	\begin{tabular}{lcccccccccr} 
		\hline
		\noalign{\smallskip}
	        \noalign{\smallskip}
		Run &  $M$ & $V_{r}$ & $V_{\phi}$ & $\delta$ & $B_r$ & $B_z$ & $B_{\phi}$ \\
		\hline
		\noalign{\smallskip}
		\textbf{HD }\\
		\hline
		\noalign{\smallskip}
		HD1 & 1 & 0 & 0 & 0.1 & - &- & -\\
		HD3 & 3 & 0 & 0 & 0.1 & - &- & -\\
		HD10 (Reference run) & 10 & 0 & 0 & 0.1 &-  &- & -\\
		 HD10t &  10 &  0 &  0 &  0.1 &-  &- & -\\
		HD25 &  25 & 0 & 0 & 0.1 & -&  -& -\\
		 HD3D &  10 &  0 &  0 &  0.1 &-  &- & -\\
		HD10Vr & 10 & 0.01 & 0 & 0.1 &-& -& -\\
		HD10Vphi & 10 & 0 & 0.01 & 0.1 & - & -& -\\
	        %
		HD-Qcl (quiescent clump) & 10 & 0 & 0 & 0.1 &  - &- & -& -\\
		HD-Ecl (explosive clump) & 10 & 0 & 0 & 0.1 & -  &- & -& -\\
		\noalign{\smallskip}
		\hline 
		\noalign{\smallskip}
		\textbf{\sn Interacting jets}\\
		\noalign{\smallskip}
		 HD10-2jet50 (2 jets, second jet at 50) &  10 &  0 &  0 &  0.1 &-  &- & -\\
		 HD10-2jet100 (2 jets, second jet at 100) &  10 & 0 &  0 &  0.1 &-  &- & -\\
		 HD10-3jet (3 jets) &  10 &  0 &  0 & 0.1 &-  &- & -\\
		\noalign{\smallskip}
		\hline 
		\noalign{\smallskip}
		\textbf{MHD }\\
		\noalign{\smallskip}
		\hline
		\noalign{\smallskip}
		MHD-Br & 10 & 0 & 0 & 0.1 &   0.001 & 0 & 0 \\
		MHD-Bz &10 & 0 & 0 & 0.1 &   0 & 0.002 & 0 \\
		MHD-Bphi &  10 & 0 & 0 & 0.1 &   0 & 0 & 0.001 \\
		MHD-Bp & 10 & 0 & 0 & 0.1   & 0.001 & 0.002 & 0 \\
		\noalign{\smallskip}
		\hline
		\noalign{\smallskip}
		\multicolumn{9}{l}{ 1) HD3D is a run in three dimensions.}\\
		\multicolumn{9}{l}{2) HD10t is a run with the transient jet which is switched off at time 500.}\\
		\label{TBL1}
		\end{tabular}
\end{table*}

%
\subsection{Numerical setup}
Our simulations are performed in axisymmetry applying cylindrical coordinates ($r,\phi,z$)
 and thus the $z$-axis is the symmetric axis.
 {We perform the 3D run of jet-gas interaction and present and discuss that run in section \ref{3D run}.}
 
The computational domain spans a rectangular grid region applying a uniform 
spacing in the radial and the vertical direction.
The domain extends to $60\times 100$ jet radii $r_{\rm jet}$ on a grid of 
$1200 \times 2000$ cells, resulting in a resolution of $\Delta r= 0.05$. 
For the boundary conditions, the axisymmetry on the rotation axis and the standard 
PLUTO outflow (zero gradient) condition for the upper z and outer r boundary are imposed.
On the lower z boundary, we implement the jet input for the jet area $(r<1)$ 
and for the rest, the standard PLUTO reflective boundary condition is applied.
It should be mentioned that in all runs presented here the injected jet is continuously powered.

For the spatial integration, we use linear reconstruction. 
In addition, we apply a second-order RK2 scheme for time evolution.
We consider TVDLF solver for all runs. 
For the magnetic field evolution, we apply the constrained transport method 
ensuring solenodality condition ,$\nabla \cdot B = 0$. 
\subsection{Units and Normalization}
We present our results in code units.
We choose units in which the density of the ambient gas is unity $\rho_a=1$
and if we assume a mean molecular density of $10^{3} cm ^{-3}$, a mean molecular mass weight of $\mu =2.1$
the gas density of the ambient is $\rho = 3.51 \times 10^{-21} {\rm gr}{\rm cm} ^{-3}$).
{ The jet density is lower $\rho_{\rm j} = \delta \rho_a$ than the ambient gas where $\delta$ denotes 
the the density contrast between the jet and the surrounding gas.}
In addition, distances are expressed in unit of the jet radius $r_{\rm j}$
which is about $100 AU$ in physical expression.
The sound speed of the ambient gas is unity $C_{\rm s}$ which is about 1 ${\rm km} s^{-1}$and
the unit of the gas velocity corresponds to the sonic Mach number.
The physical time unit in our setup is about 500 yr.

\subsection{PROBABILITY DENSITY FUNCTION (PDF)}
The turbulence in the literature is known as the random motions of the gas flows
at different scales. To investigate the turbulence induced by the stellar jet it is important to use a proper method to
estimate the ``induced turbulence'' in the surrounding gas materials.
One of the best way to measure the turbulent motion in a flow is using the probability density 
function of velocity which has been used by previous authors (see e.g. \cite{2007ApJ...668.1028B}).
To calculate the probability density function, we consider all the volume excited by the jet propagation having the 
non-zero velocity i.e., $v_p=\sqrt{{v_r}^2+{v_z}^2}$.
We consider different velocity bins,i.e., $v_i, v_i+dv$ and the probability density function is defined as,

\begin{equation}
P_i(f_i)= w(f_i)/N_{tot} ;
\end{equation}

Where $f_i$ is a function of velocity determining the given velocity in a range of ($v_i, v_i+dv$). 
$w$ counts the number of cells that the condition  $f_i$ is valid for them.
$N_{tot}$ is the total number of cells in the computational domain. 
Thus, $P_i(f_i)$ is the probability of the condition $f_i$ in the whole domain.
The total probability used to normalize the PDF is,

\begin{equation}
\sum_i P_i = 1.
\end{equation}
\section{RESULTS AND DISCUSSIONS}
\label{sec:analysis} 
{To investigate the efficiency of driving the turbulence by the stellar jets into the ambient gas, 
We present a case study of different parameters runs including the jet Mach number, the jet velocity field 
and the background magnetic field geometries and the interacting jets}.
Also, we discuss the environmental effects on the jet-gas interaction by considering
the non-homogeneous surrounding gas containing the clumps in the model setup (see Table \ref{TBL1}).
\footnote{To visualize the results we use Python 2.7.13 and Visit 2.13.2 software.}

\begin{figure*}
\centering
\includegraphics[width=16cm]{\figurepath/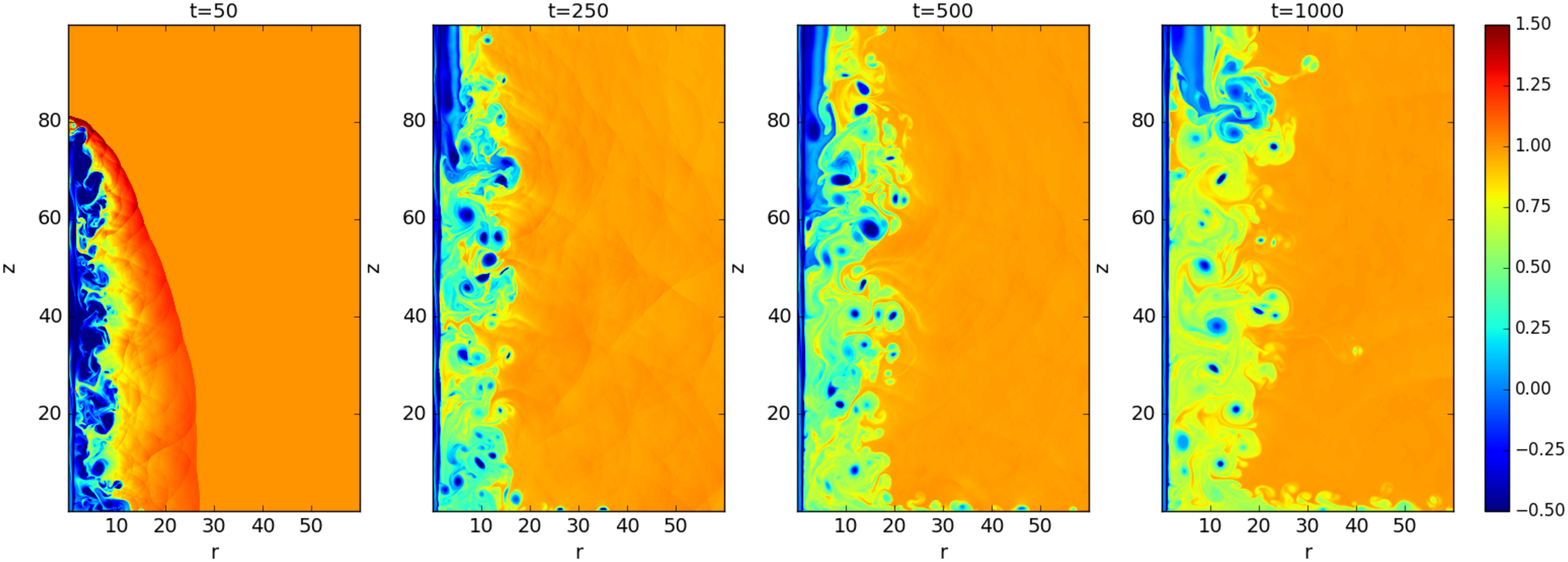}
  \caption{Reference run. Shown are the snapshots of the
  mass density in Logarithm scale for reference run {\em HD10} at times 50, 250, 500, 1000.}
 \label{fig:ref_den}
 \includegraphics[width=16cm]{\figurepath/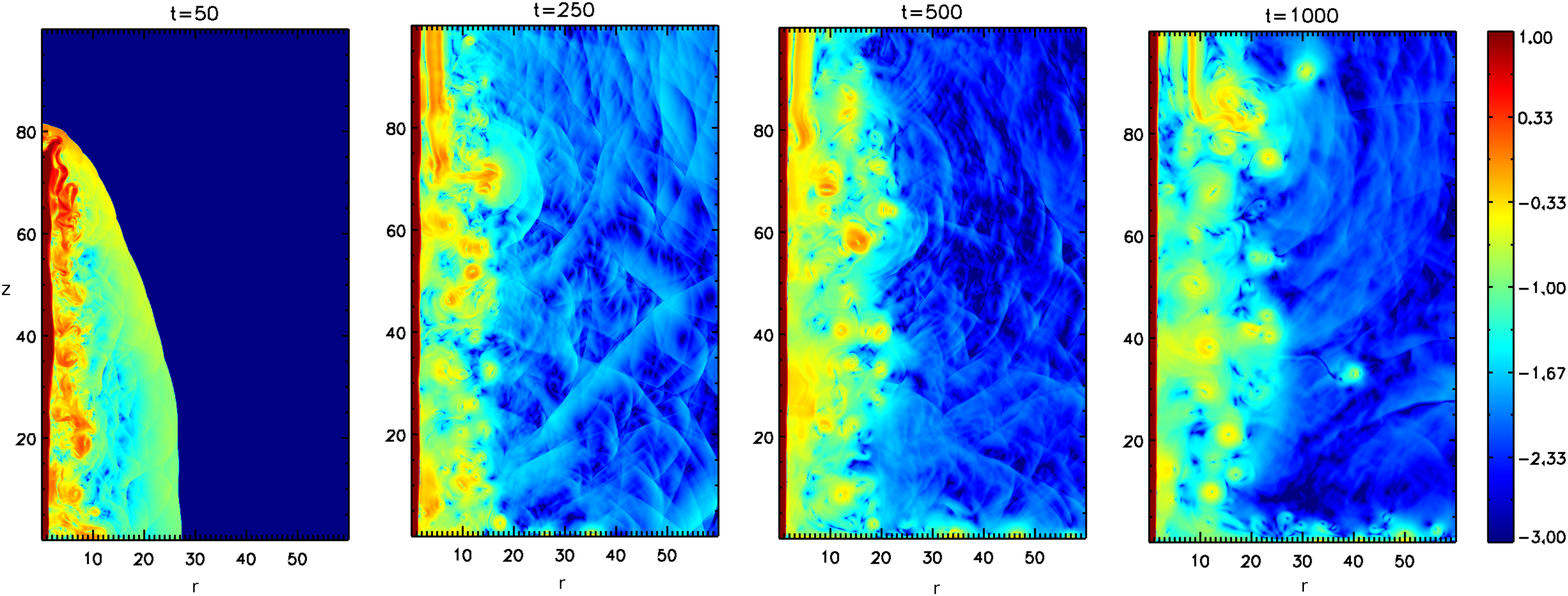}
  \caption{Shown are the snapshots of the poloidal velocity in Logarithm scale, $v_p=\sqrt{v_{r}^2+v_{z}^2}$, for reference run {\em HD10}
  at times 50 ,250, 500, 1000.}
  \label{fig:ref_logV}
  \includegraphics[width=16cm]{\figurepath/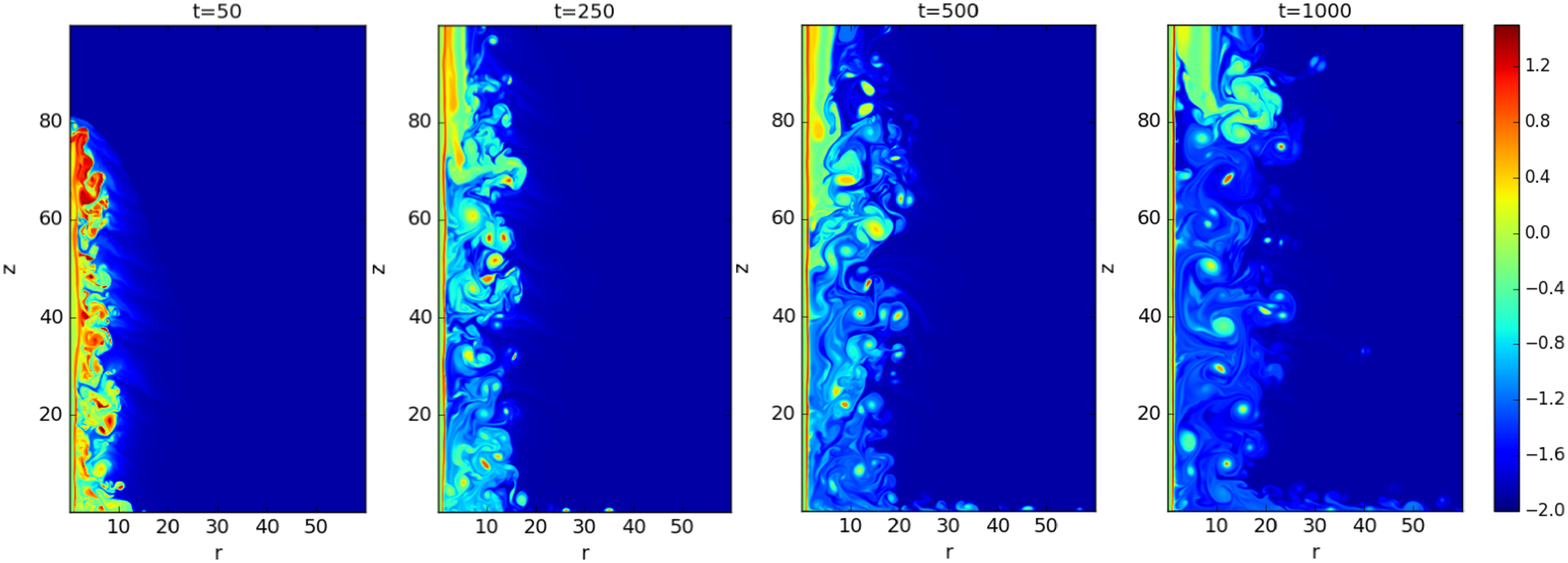}
  \caption{Displayed are the snapshots of the entropy of reference run {\em HD10} at times 50, 250, 500, 1000.}
  \label{fig:ref_entropy}
  \end{figure*}
\begin{figure*}  
\centering
\includegraphics[width=16cm]{\figurepath/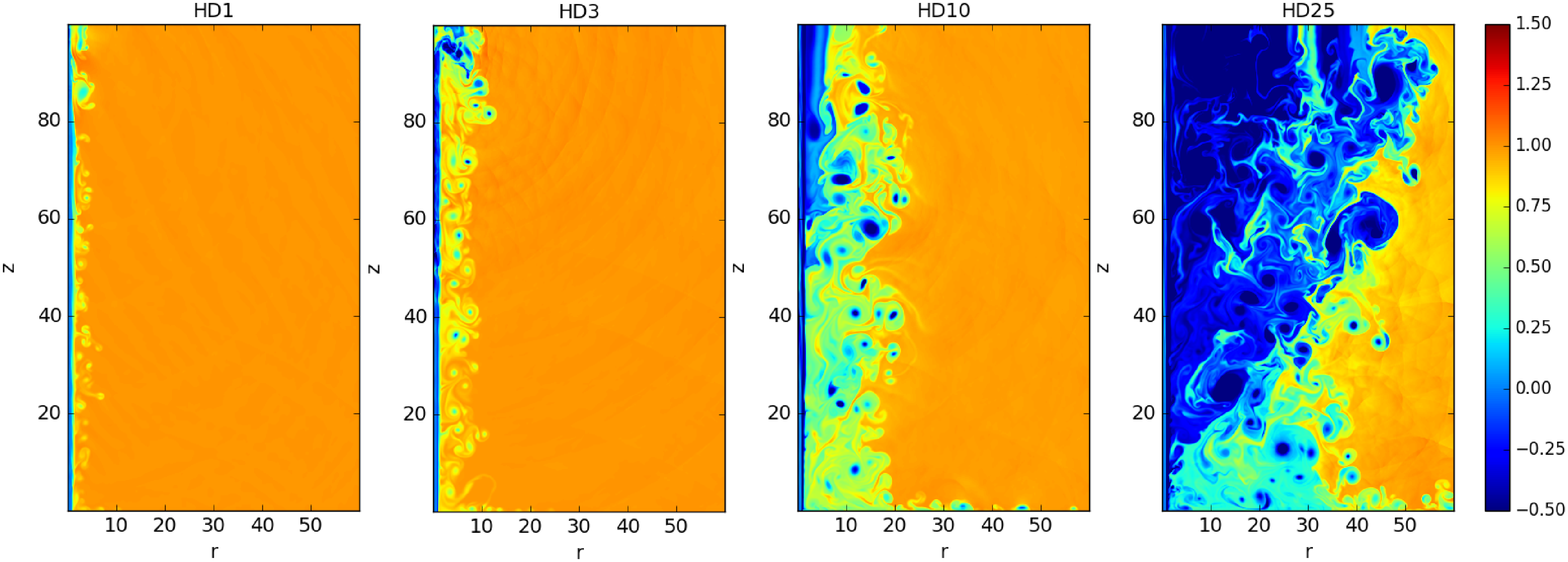}
  \caption{Shown are the snapshots of the mass density in Logarithm scale for
  runs with Mach numbers of 1, 3, 10 and 25 at time 500.}
\label{fig:compare_den_Mach}
\includegraphics[width=16cm]{\figurepath/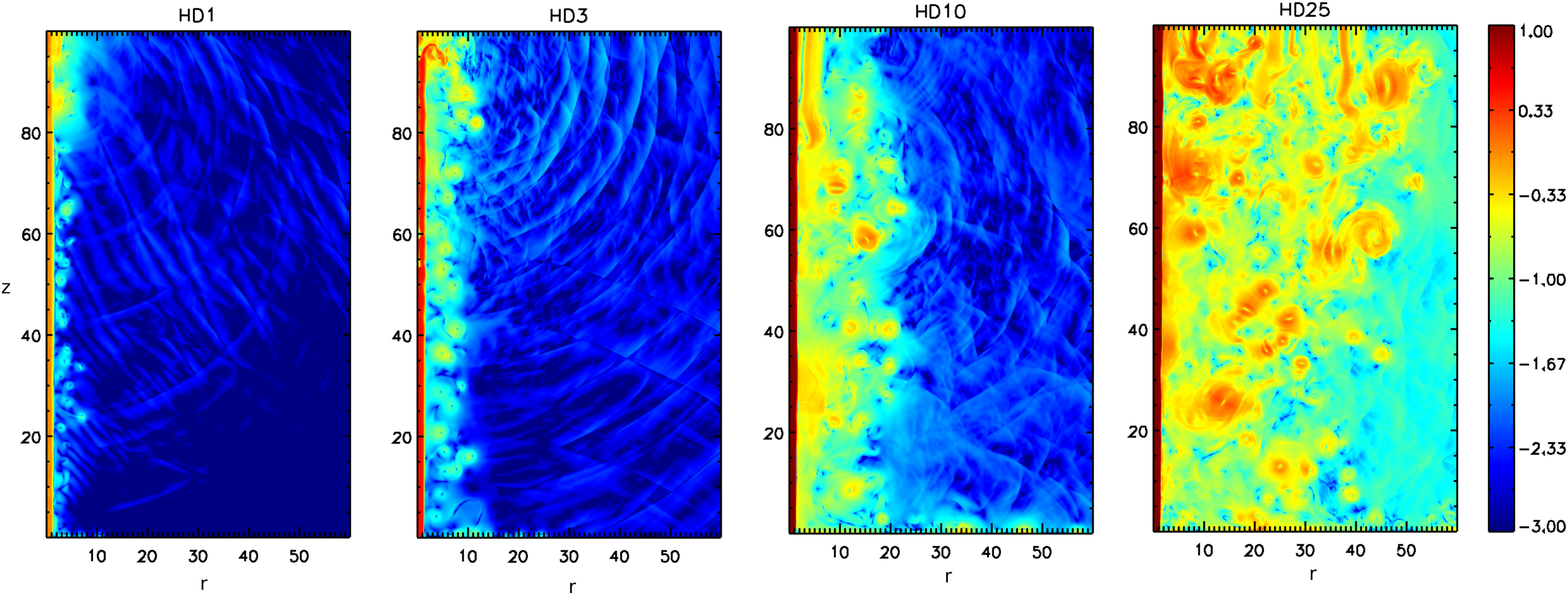}
  \caption{Shown are the snapshots of the velocity field in Logarithm scale $v_p=\sqrt{v_{r}^2+v_{z}^2}$ 
  for runs with Mach numbers of 1, 3, 10 and 25 at time 500}
\label{fig:logV_power}
\includegraphics[width=16cm]{\figurepath/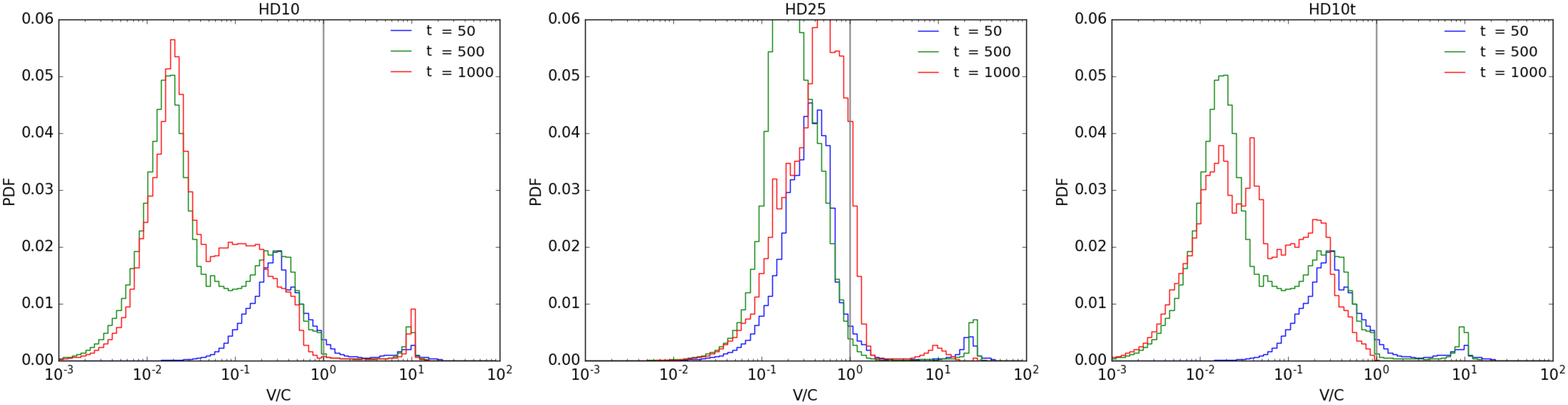}
  \caption{ Shown are the plots of the probability density function of velocity for run {\em HD10},{\em HD25} 
 and  {\em HD10t} at times 50, 500, 1000. Here C denotes to the local sound speed of the gas.
 The vertical line shows the transonic velocity and distinguishes between the subsonic and the supersonic velocities.
 }
\label{fig:compare_pdf}
\end{figure*}
\begin{figure}
\centering
\includegraphics[width=1.0\columnwidth]{\figurepath/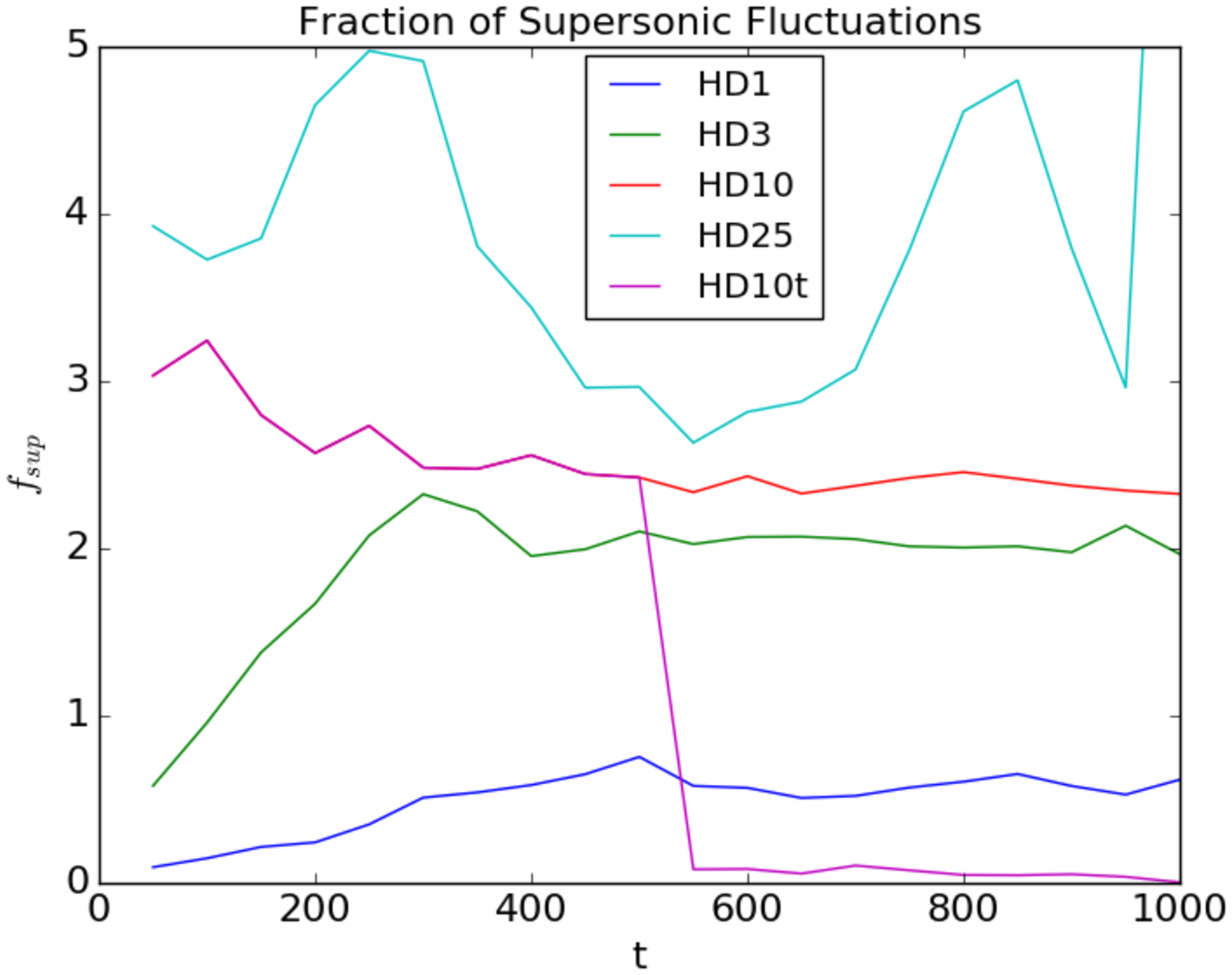}
  \caption{ Shown is a comparison of the evolution of supersonic features with respect to all gas  materials
  for runs with Mach numbers of 1, 3, 10 , 25 and run with the transient jet. }
\label{fig:Sup_fraction_powers}
\end{figure}
%
%
{  First, we consider the effects of jet Mach number on the jet-gas interaction.}
As shown in Table \ref{TBL1} we perform runs with different Mach numbers covering the transonic to highly supersonic jet .
Figure \ref{fig:compare_den_Mach} displays the snapshots of the mass density at late evolutionary stages for runs {\em HD1, HD3, HD10,} and {\em HD25}
with Mach numbers of 1, 3, 10 and 25, respectively.
An obvious correlation between the jet Mach number and the size of the excited gas is found. 
In particular, it is observed that the more random motions and more filamentary structures are produced by the jet
with the higher Mach number. 

{ Another useful parameter is the entropy of the gas displaying the energy distribution of the 
system and is shown in Fig \ref{fig:ref_entropy}. 
It demonstrates that the jet materials have higher entropy than the entrained gas.
As time passes, the excited gas loses the energy, gets slower and reaches a lower entropy level. }

We can trace the evolution of the jet and the surrounding gas by considering the probability density 
function of the velocity, more precisely.
To calculate the PDF, we consider the local sound speed $c=\sqrt{(\gamma p/\rho)}$ of the gas which allows for distinguishing the 
supersonic and the subsonic features.

{
 In order to investigate how long the velocity fluctuations produced by the stellar jet can survive in the environment
we perform a run including the transient jet  called {\em HD10t} in which the jet engine is switched off after time 500  and considered in the PDF diagram as well.

Figure \ref{fig:compare_pdf} illustrates the PDF of velocity for reference run, run {\em HD25} and run {\em HD10t} with the transient jet
at times 50, 500, 1000. 
It is clearly seen that the transient jet (right panel) does not show any significant supersonic fluctuations after the driving has been stopped.
In fact the dying jet displays the decay of the peak of the PDF of velocity after switching off the jet  and no super sonic feature is seen after time 500.

Comparing two other runs with Mach number 10 and 25, we find that in both runs, the small fraction of the excited gas is supersonic
and the majority of the motions are subsonic.
But, the peak of the PDF of velocity stays at higher velocity in run with higher Mach number, which means that
the entrained gas has a higher velocity.
Also, the velocity map shown in Fig \ref{fig:logV_power} confirms that the entrained gas extends to the larger area 
and corresponds to the higher velocity in a run with a higher jet Mach number.
By passing time, the peak of the PDF of velocity moves towards the lower velocities which indicates the decay of the 
induced turbulence in the entrained gas. 
In comparison, we do not see a big variation in the position of the peak of the PDF of velocity in run {\em HD25} with 
the higher Mach number during the time.
}

Specifically, it is possible to compare the supersonic part of the gas of all runs and have a better view.
{ Figure \ref{fig:Sup_fraction_powers} illustrates the evolution of the supersonic features with respect to all gas materials for
runs applying different Mach numbers and run with the transient jet.}
It shows that some part of the entrained gas remains supersonic at the late evolutionary stage (t=1000).
Since the jet with the higher Mach number interacts more strongly and also transfers a larger fraction of energy and angular momentum 
the size of the excited gas increases and the peak of the PDF of velocity fluctuations stays at larger velocities.
It means that the velocity of the injected jet can affect the distribution of the velocity fluctuations and 
the induced turbulence.
Thus, one can conclude that the supersonic turbulence driven by a jet with higher Mach number 
can survive on a larger time scale.
This result is different from \citet{2007ApJ...668.1028B}. 
They discuss that a slower jet produces a larger fraction of excited gas.
It should be noticed that our study corresponds to a different length scale and time scale.
Our analysis is constrained to the box size which is about $10^4$ AUs and
is smaller.
Also, the time scale in our study is smaller and the surrounding gas is still affected by jet propagation in our simulations.
In addition, we  discuss the turbulence  in general and we are not constrained to the supersonic turbulence 
which is different from other studies.

\subsection{Impact of the jet velocity field}
Besides the jet Mach number, it is also important to see how different velocity components 
of the jet materials can influence the jet-ambient gas interaction.
To have better comparison and rather wider case study compared to previous works  \citep{2007ApJ...668.1028B},
we perform runs with different initial velocity field of the jet area.
Here, we present two runs {\em HD10Vr} and {\em HD10Vphi} including the radial and the rotation velocity together with the vertical velocity.

The evolution of their mass density are shown in Fig \ref{fig:compare_Vr}.
Compared to the reference run including the jet radial velocity or the jet rotation increases the size of the entrained gas.
This behavior is seen more clearly in the PDF of these two runs (see  Fig \ref{fig:compare_pdf_Vr_Vphi}).
In comparison, we observe that the runs with additional velocity than $v_z$ have a wider PDF at the late evolutionary stage.
It means that the entrained gas covers a wider range of velocities and the position of the peak of the PDF of velocity stays at the 
larger velocity.
In addition, we find that the peak of the subsonic features in a run applying the rotating jet
stays at higher velocities and the stronger effects on the surrounding gas are found.
This behavior can be explained by considering the centrifugal force of different runs.
Since the rotation is initially chosen to be zero in other runs, the centrifugal force appears only 
in a run with the rotating jet {\em HD10Vphi}  and the shearing motions between the jet and the ambient gas increase.
As we mentioned in the model setup the jet is in pressure equilibrium with the surrounding gas initially.
Thus, the jet is stable itself and it interacts by shear instabilities with the surrounding gas.
Consequently, we find that the jet rotation enhances the shearing motions which results in
turbulent fluctuations in the ambient gas.
%
\begin{figure}
\centering
\includegraphics[width=1.0\columnwidth]{\figurepath/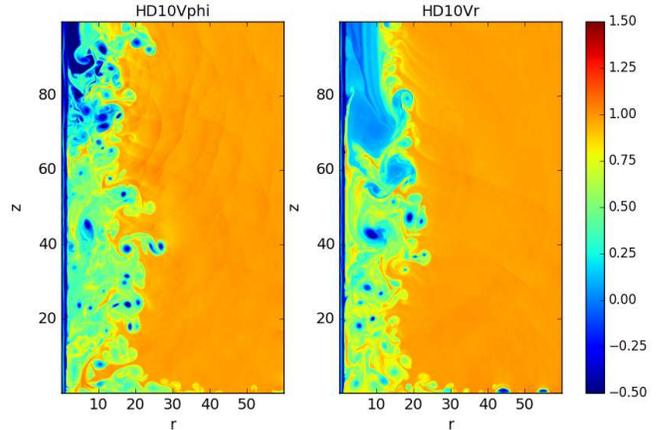}
  \caption{Shown are the snapshots of the
  mass density in Logarithm scale for runs {\em HD10Vr}, {\em HD10Vphi} including radial
  and rotational jet velocity, respectively. }
\label{fig:compare_Vr}
\end{figure}

\begin{figure*}
\centering
\includegraphics[width=16cm]{\figurepath/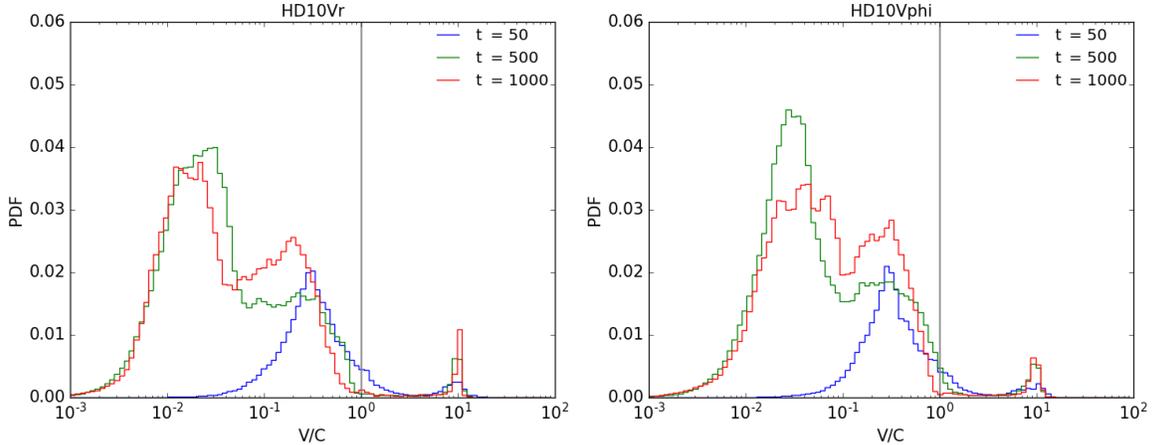}
  \caption{Shown are the PDF of velocity for runs {\em HD10Vr}, {\em HD10Vphi}, at times $t=50$, $t=500$, $t=1000$.}
\label{fig:compare_pdf_Vr_Vphi}
\end{figure*}

\begin{figure*}
\centering
\includegraphics[width=18cm]{\figurepath/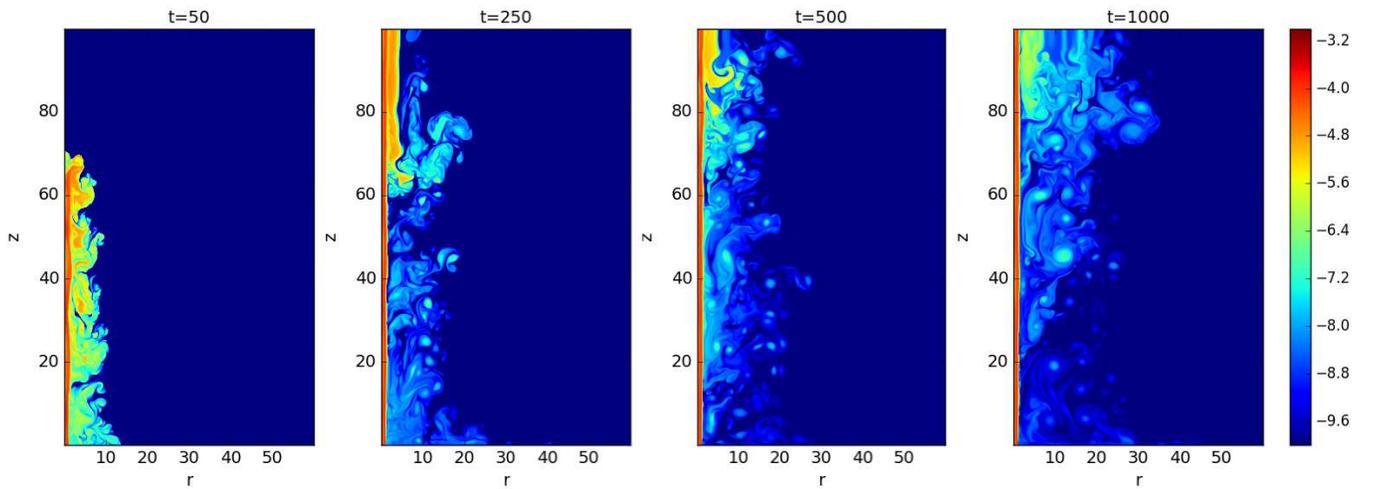}
  \caption{Centrifugal force. Shown are the snapshots of the centrifugal force for run {\em HD10Vphi}  including jet rotational velocity. }
\label{fig:Fcen_Vphi}
\end{figure*}


\subsection{Magnetized jet-ambient gas interaction}
According to the observational evidence the magnetic field exists in the jet source, inside the jet 
\citep{1984RvMP...56..255B, 1999ApJ...520..706C, 2001ApJ...546..980O, 2008Natur.452..966M, 2016ApJ...817...96G}
and also in the interstellar gas \citep{1999ApJ...514L.121C,2017A&A...601A..90B, 2018A&A...614A.100T}.
It is also believed that jets are most probably magnetically launched and collimated 
\citep{1982MNRAS.199..883B, 1983ApJ...274..677P, 1997A&A...319..340F, 2007prpl.conf..277P}.

Here, we discuss the effects of the background magnetic field on the jet-ambient gas interaction.
We present the MHD simulations applying different background magnetic field geometries, i.e., run {\em MHD-Br} applying the radial magnetic field
,{\em MHD-Bz} applying the vertical magnetic, run {\em MHD-Bphi} applying the toroidal magnetic field and run {\em MHD-Bp} applying the poloidal 
magnetic field in the domain (see Table \ref{TBL1}).
%

\begin{figure*}
\centering
\includegraphics[width=18cm]{\figurepath/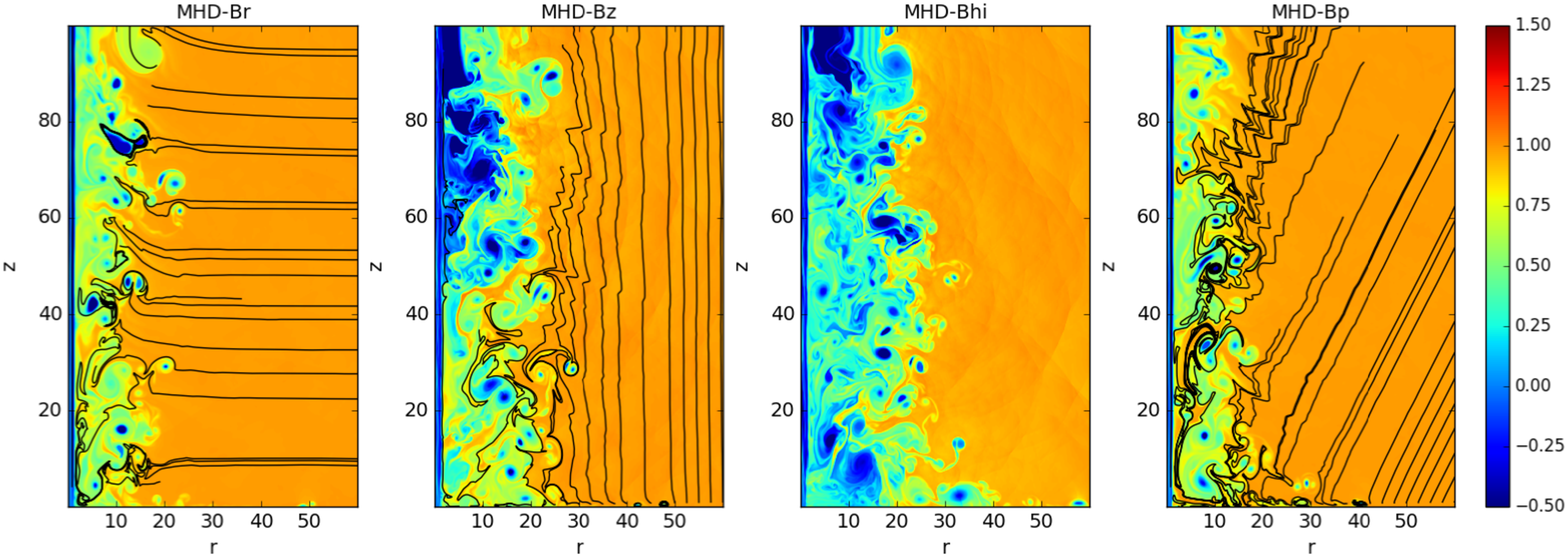}
  \caption{ Shown are the snapshots of the mass density in Logarithm scale for MHD runs applying different background magnetic fields at time 500.
  The field lines are shown in black.}
\label{fig:compare_den_B}
\centering
\includegraphics[width=18cm]{\figurepath/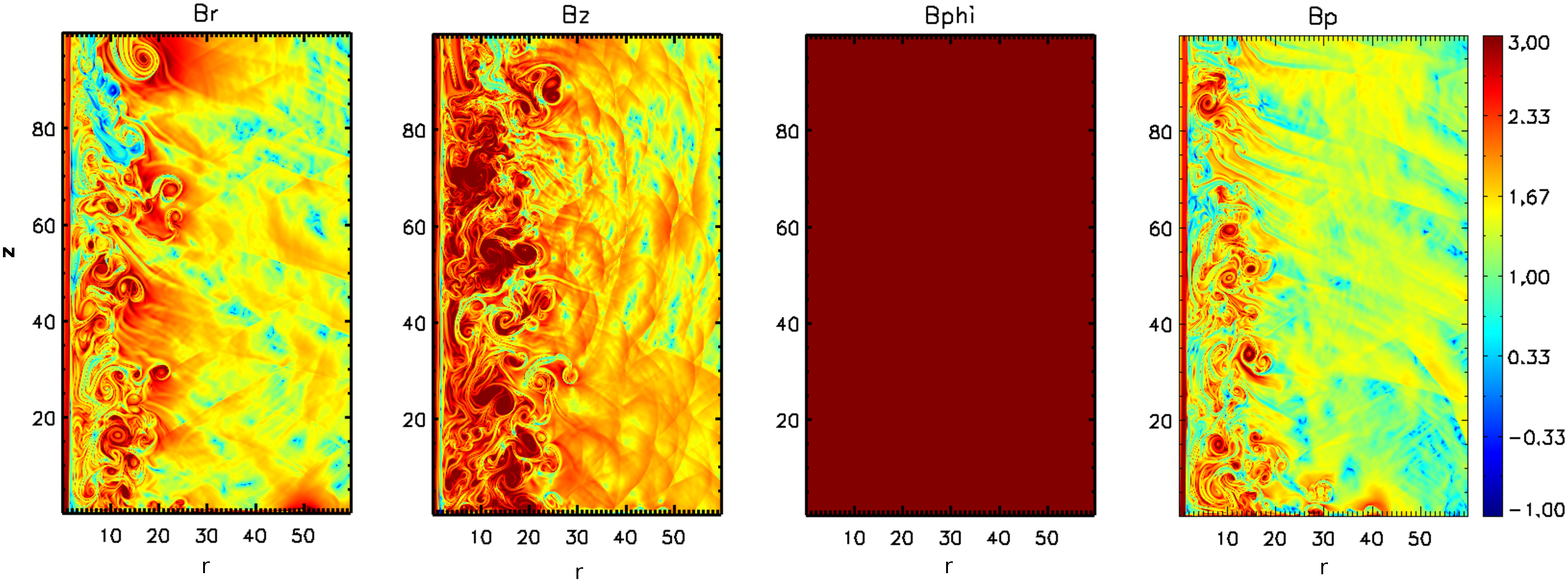}
\caption{ Shown are the snapshots of the Alfven Mach number $M_A=v_p/v_A$ in Logarithm scale for
MHD runs applying different background magnetic fields at time 500.}
\label{fig:Alfven_MHD}
\centering
\includegraphics[width=18cm]{\figurepath/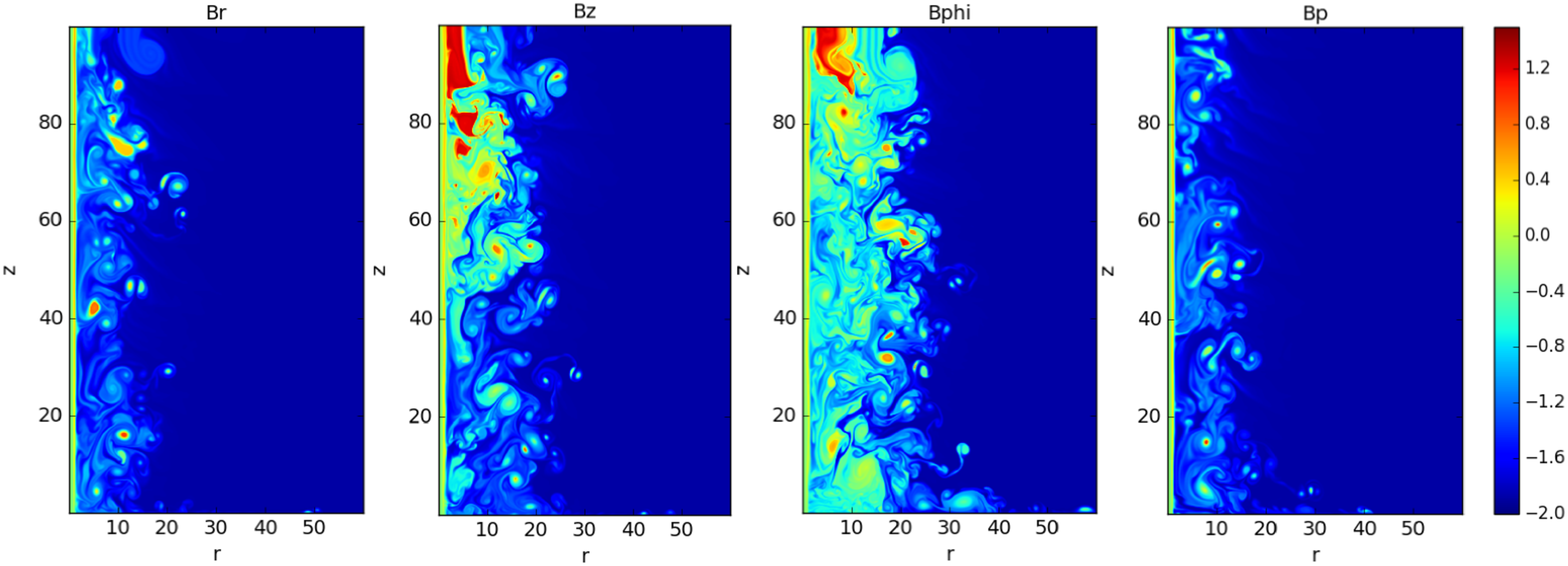}
  \caption{Shown are the snapshots of the entropy of gas in Logarithm scale for MHD run applying different background magnetic fields at time 500}
\label{fig:s_MHD}
    \centering
\end{figure*}
 
\begin{figure*}
 \centering
    \includegraphics[width=14cm]{\figurepath/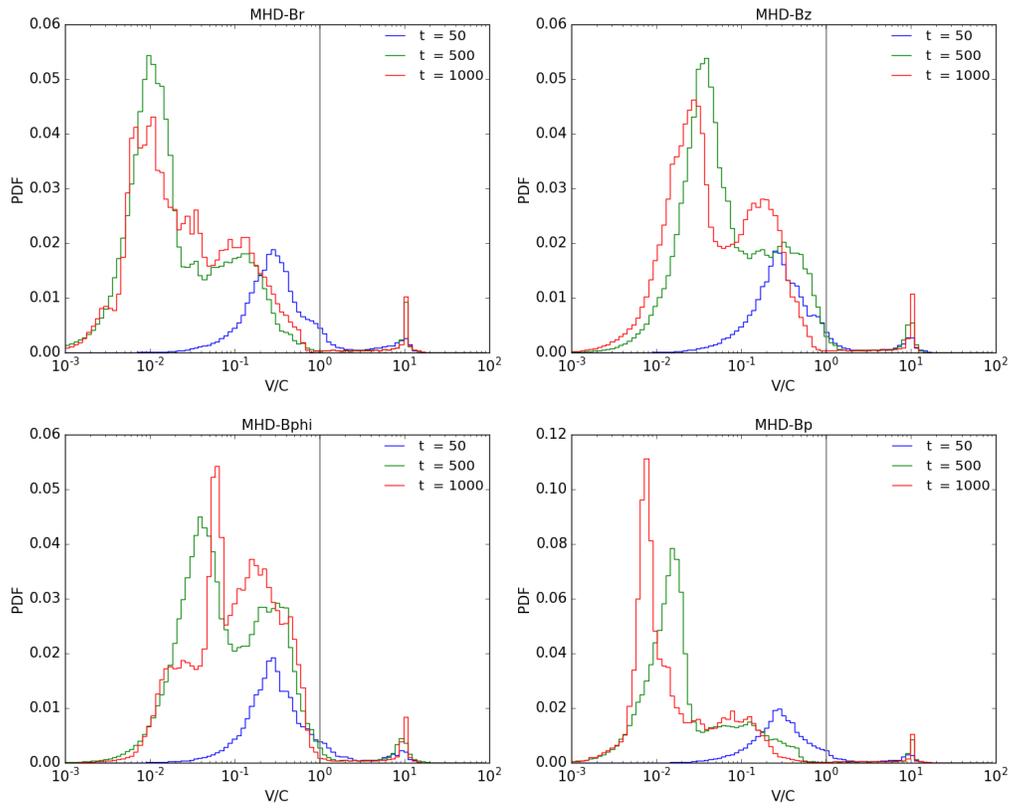}
    \caption{Shown are the Probability density functions of velocity for MHD runs at different times.}
    \label{fig:MHD_PDF}
\end{figure*}
 \begin{figure}
\centering
\includegraphics[width=1.0\columnwidth]{\figurepath/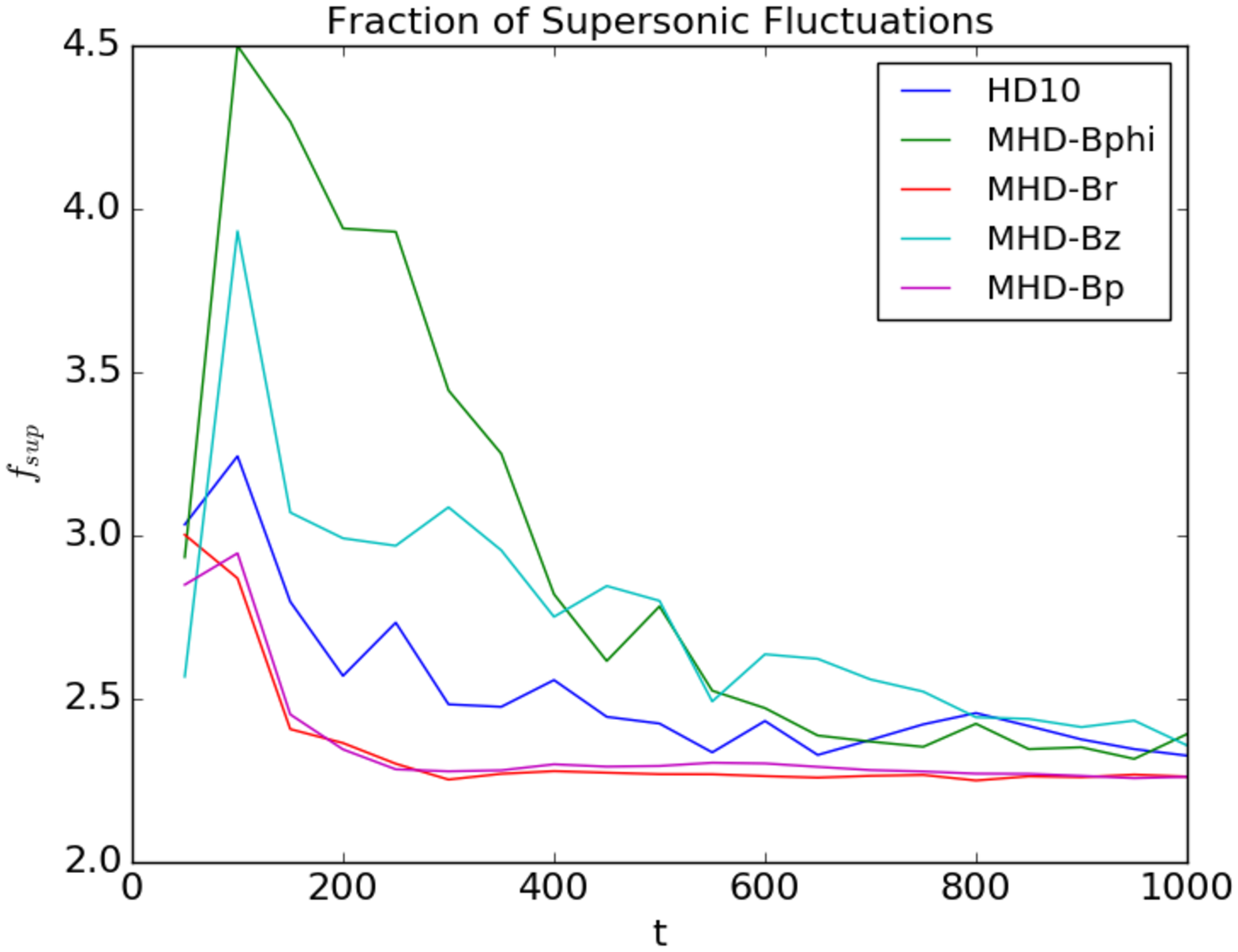}
  \caption{ Shown is a comparison of the evolution of the supersonic features with respect to all gas materials
  for the MHD runs and the reference run.}
\label{fig:Sup_fraction_MHD}
\end{figure}

For the sake of comparison, we consider different physical parameters of all MHD runs.
First of all, we consider the mass density distribution of all MHD runs.
Figure  \ref{fig:compare_den_B} displays the snapshots of the mass density in Logarithm scale for
MHD runs applying different background magnetic fields at time 500.
  The field lines are shown in black. It is seen that the final structure of the entrained gas 
depends on the  background magnetic field distribution.
We observe that a run with the radial or poloidal background magnetic field produces the denser filamentary structures
and the entrained gas is confined to a smaller area.

 The excited gas materials and the  filamentary structure and random motions are seen very nicely in Alfven Mach number maps.
Figure \ref{fig:Alfven_MHD} displays the snapshots of the Alfven Mach number in Logarithm scale for all MHD runs.
Here the Alfven velocity and the Alfven Mach number are $v_A=\sqrt{B_p^2/4\pi\rho}$, $M_A=v_p/v_A$.
We observe that the entrained gas is strongly super Alfvenic in all MHD runs that indicates the Alfven velocity is not large in our MHD setups.
It is clearly seen that in run with the poloidal field (last panel) the Alfven Mach number is lower.

In addition, the entropy maps of all MHD runs shown in Fig\ref{fig:s_MHD}.
We observe that in a run with the toroidal or vertical magnetic field a larger amount of energy 
is transferred into the ambient gas.
It is consistent with the fact that the larger volume of entrained gas is produced and 
more fluctuations are seen in run with the toroidal or vertical magnetic field.
It seems that the magnetic Lorentz force plays an important role and thus we observe different degree of 
collimation or different accelerations around the jet area.

Considering the PDF of velocity of all MHD runs shown in Fig \ref{fig:MHD_PDF}, it is observed that in 
a run with the radial and poloidal magnetic field the excited gas materials are slower.
This behavior is consistent with the mass density pattern which contains the denser features in the entrained gas (see Fig \ref{fig:compare_den_B}).
In comparison, the bulk of the subsonic turbulent features in the run with the toroidal field 
stays at the larger velocities which is consistent with the higher entropy level in this run.

In order to be able to compare the supersonic turbulence, we provide Fig \ref{fig:Sup_fraction_MHD}
 showing the fraction of the domain having supersonic velocities, for all MHD runs. 
In fact, in all runs a small fraction of supersonic fluctuations is produced in the interstellar gas which is decreasing during the time.
However, it is seen that in a run with the toroidal field, the fraction of the supersonic features in the ambient gas is 
higher than reference run and other MHD runs.
One physical reason for this behavior is that the run with the radial or the poloidal magnetic field look 
more collimated and thus the interaction between the jet and the ambient gas is less.
Consequently, the shearing motions between the jet and the surrounding gas are less.
There should be also other physical reasons which are not known completely and a deeper study is needed.

In total, we find that the run with the toroidal back ground magnetic field has larger impacts on ambient gas and
provides more efficient circumstances to transfer and maintain the turbulence in the surrounding gas.
It should be noticed that the entrained gas is almost subsonic and a small fraction of that stays in supersonic regime.

\subsection{Jet propagation in the clumpy environment}
Observational studies demonstrate the interstellar gas is not necessarily homogeneous 
and jets may encounter clumpy regions during their propagation in the surrounding 
gas \citep{2007ApJ...654..304K, 2009A&A...504..415S}.

%
\begin{figure*}
\centering
\includegraphics[width=18cm]{\figurepath/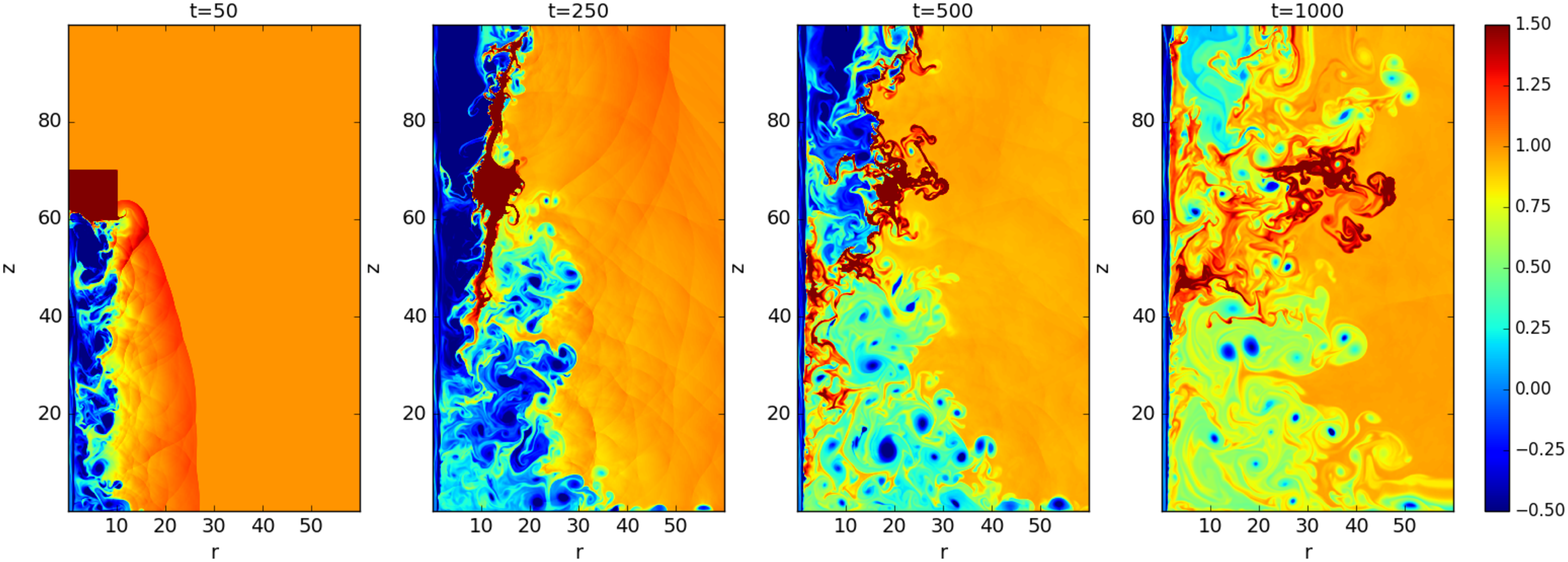}
\includegraphics[width=18cm]{\figurepath/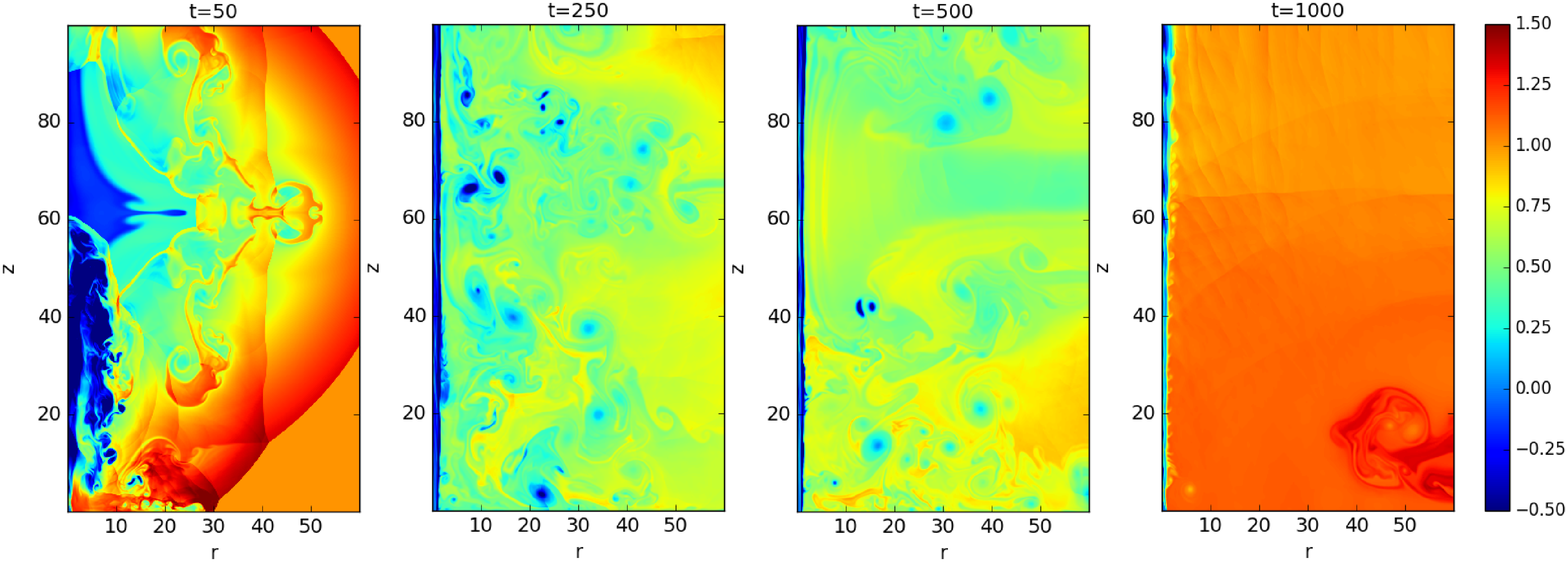}
  \caption{Shown are the snapshots of the mass density for run {\em HD-Qcl} (top) and run {\em HD-Ecl}
  at times 50,250, 500, and 1000 .}
\label{fig:ref_denclump}
\end{figure*}
\begin{figure*}
\centering
\includegraphics[width=15cm]{\figurepath/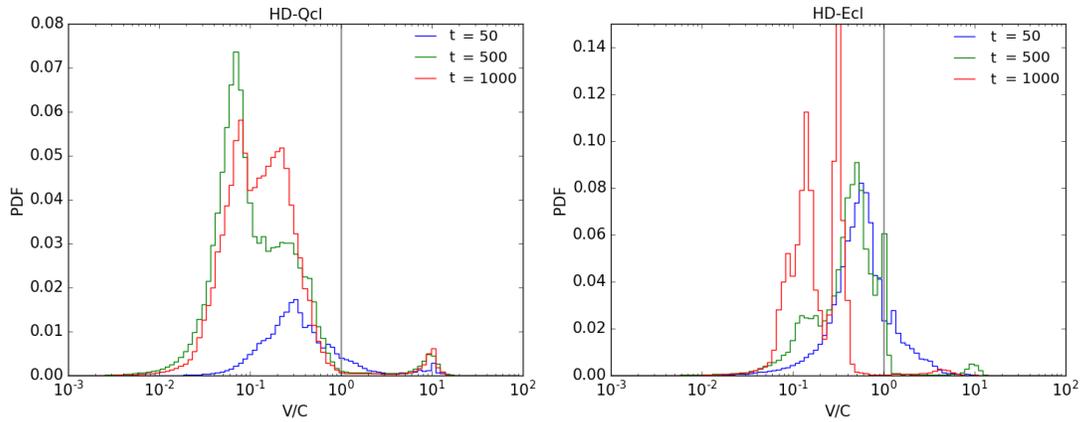}
  \caption{Shown are the Probability density functions of velocity for runs, {\em HD-Qcl} and {\em HD-Ecl} at different times.}
\label{fig:clumpPDF}
\end{figure*}

In this section, we present the results of simulations in which the jet is propagating into the non-homogeneous interstellar gas 
and interacts with the clump.
We use two different methods to define the clump region.

In the first approach, we define the clump as a region with the higher density but in
pressure equilibrium with its neighboring gas, i.e., run {\em HD-M10cl} and we call it ``quiescent clump''.
In the second approach, the clump region is defined with a higher density and higher
pressure than the neighboring gas, i.e., run {\em HD-M10pcl} which is called ``explosive clump''(see Table \ref{TBL1}).

Figure \ref{fig:ref_denclump} displays the snapshots of the mass density at different times for 
these two runs.
It is observed that the ``quiescent clump'' exists for a longer time scale and 
disturbs the jet area locally.
Instead the ``explosive clump'' appears temporary and acts like a violent 
perturbation and injects energy and disturbs the environment strongly.
We recognize that by applying the non homogeneous ambient gas the more entrained gas and 
random motions are produced compared to the reference run.

If we compare the PDF of velocity of two clumpy runs we recognize that a considerable fraction of
supersonic features are produced, initially in run containing the ``explosive clump'' (see Fig \ref{fig:clumpPDF}). 
However, the supersonic fluctuations are not survived until the late evolutionary stage (t=1000).
It is clearly seen that the peak of the velocity fluctuations moves toward the smaller values 
reflecting that the turbulence decays rather quickly in this run.
It seems that the interaction of the jet and the surrounding gas, in the explosive run is quick and strong but not so efficient to 
keep the turbulence in the environment for a long time.

As a result, we find that the environmental effects are important in the evolution and maintenance of the turbulent motion
in the interstellar gas.
In particular, the run containing the long lasting clump ``quiescent clump'' provides the more efficient circumstances
in which the driven turbulence can be maintained for a longer time scale.

\subsection{Interacting Jets}
It is known that the stars are born in cluster environments 
and thus the stellar outflows might influence the cluster evolution and thus
the star formation rate and the mass spectrum within the cluster environments
\citep{2006ApJ...640L.187L, 2010ApJ...709...27W, 2014ApJ...790..128F}.

However it is still not clear whether the stellar outflows are the main driver of turbulence.
To study the efficiency of the stellar outflow as a turbulence driver on a larger scale, we perform some simulations of 
interacting jets defined in a larger computational domain.

It should be noticed that the more appropriate setup would be a computational domain defined in
Cartesian coordinates including two slab jets sufficiently
far from each other.
However, we perform the simulations of  interaction jets in the same cylindrical coordinates to have a better comparison with
our reference run.

These simulations are in hydrodynamic regime and shown in Table \ref{TBL1}, i.e., run {\em HD10-2jet50}, {\em HD10-2jet100} and {\em HD10-3jet}
including two and three interacting jets propagating in the domain.

The snapshots of the mass density of runs including the interacting jets are shown in Fig \ref{fig:rho_interact2jet} and \ref{fig:rho_interact3jet}.
It is observed that each jet propagates and produces the shocked area and the entrained materials.
By passing time, the interaction between different jets happens.
In run {\em HD10-2jet50}, the interaction happens at earlier time since the separation between jets is smaller.

The interaction between jets leaves a dense region between each two
jets and is maintained for a long time.
It should be noticed that the jet materials lose energy during the interaction and get slower.
During the interaction the jet areas are disturbed and at late evolutionary stage it seem that the jets are suppressed.
Thus, not the typical outflows with the large velocity are seen at late evolutionary stages in the environment.
Similar behavior is seen in run with three interacting jets (Fig \ref{fig:rho_interact3jet}).
It is seen that two dense regions and the large amount of entrained materials are produced in this run.

Figure \ref{fig:pdf_interacting} illustrates the probability density function of velocity distribution of runs including interacting jets.
Regarding the PDF of velocity, we find that the supersonic features are not survived for a long time.
However, comparing to the reference run the peak of the PDF of velocity is shifted toward the transonic velocities.
Specifically, it is seen  that the entrained  materials have larger velocities in run with interacting jets compared to the reference run with a single jet.
The amount of transonic gas in run with three interacting jets enhances.

As a results, we find that including the multiple jets can enhance the amount of fluctuations in the environments which are mainly in subsonic and transonic regime.
The supersonic turbulence are suppressed and can not be maintained for a long time scale after the interaction between jets happens.

\begin{figure*}
\centering
\includegraphics[width=18cm]{\figurepath/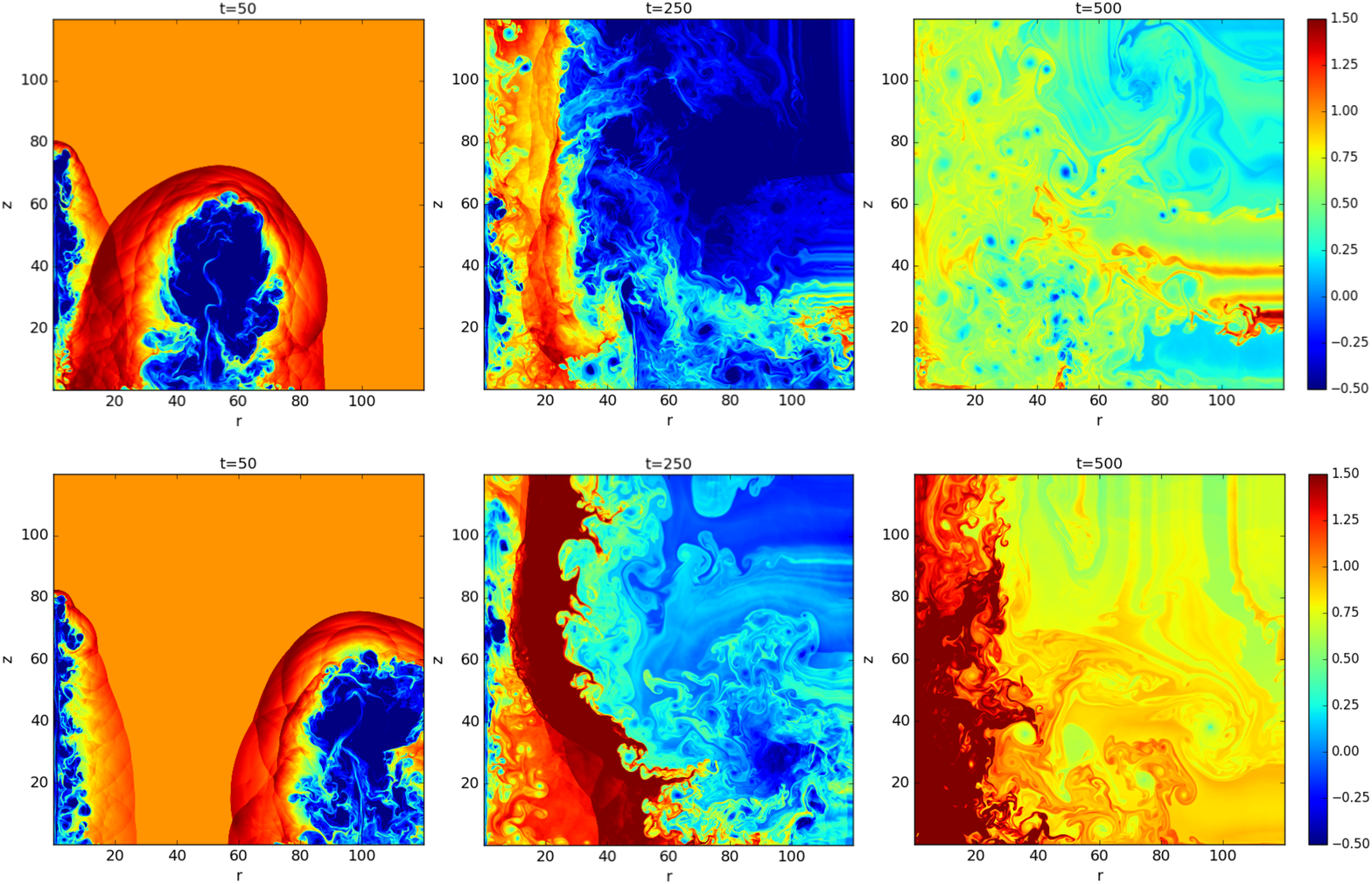}
  \caption{  Interacting jets. Shown are the snapshots of the  density mass of runs including the interacting jets, i.e.,
  run {\em HD10-2jet50} with (top) and {\em HD10-2jet100} (middle) with the position of the second jet at 50 and 100 at times 50, 250, 500.}
  \label{fig:rho_interact2jet}
  \includegraphics[width=18cm]{\figurepath/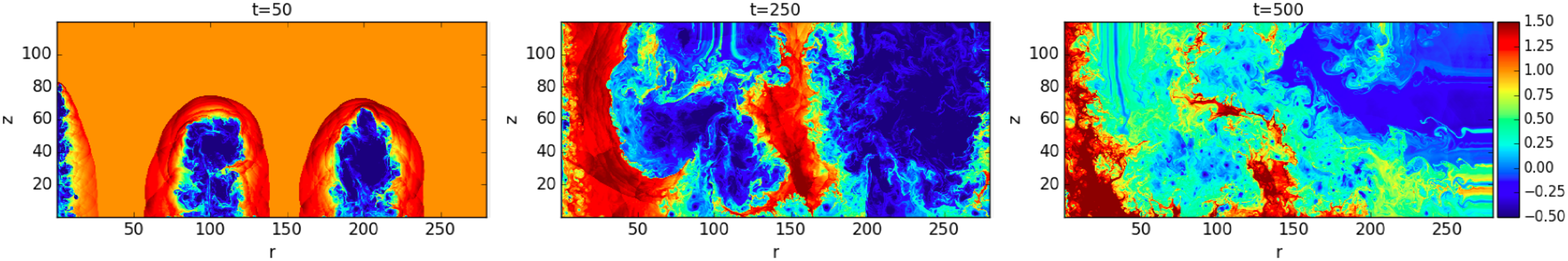}
\caption{  Shown are the snapshots of the  density mass of run {\em HD10-3jet} including  three interacting jets at times 50, 250, 500.}
\label{fig:rho_interact3jet}
\end{figure*}

\begin{figure*}
\centering
\includegraphics[width=18cm]{\figurepath/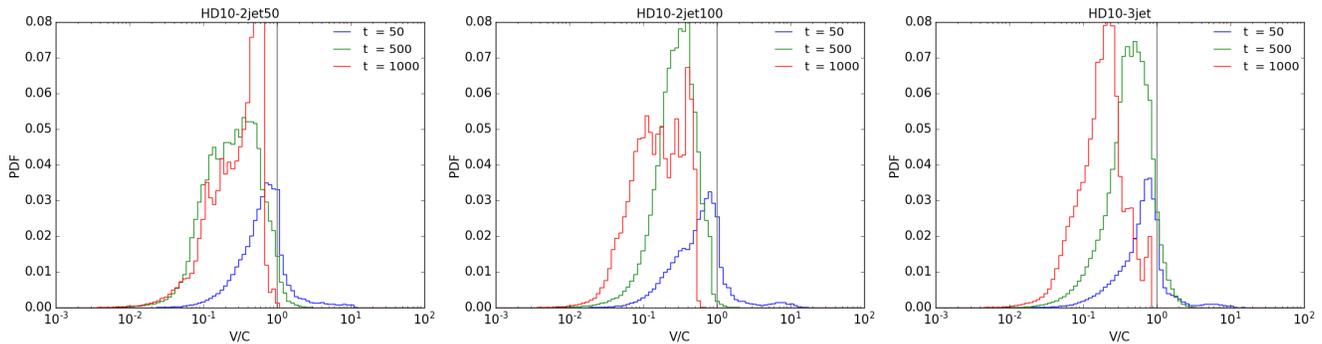}
  \caption{ Shown are the plots of the probability Density Function of velocity for run {\em HD10-2jet50}, {\em HD10-2jet100}
 and  {\em HD10-3jet} including interacting jets at times 50, 500, 1000.}
\label{fig:pdf_interacting}
\end{figure*}

\subsection{Jet-ambient gas interaction in 3D}
\label{3D run}

Observations show that the nature does not obey axisymmetry and to have a more natural view of jet-ambient gas interaction
we also perform a 3D hydrodynamic simulation, {\em HD3D} and present the results in this section (see Table \ref{TBL1}) .

In particular, we extend the reference model setup to three dimensions and apply similar initial conditions as axisymmetric run.
The computational domain is defined in polar coordinates $(r, \phi, z)$ with the z-axis chosen along the direction of jet rotation axis.
The boundary conditions and units are similar to the reference run.

The global evolution of the jet in 3D run is similar to the jet evolution in 2D reference run.
The cuts of density mass distribution in three dimensions are shown in Fig \ref{fig:run3D_rho} at different times.
The jet forms the bow shock and excites the ambient gas materials and transfers the energy and turbulence into the surrounding gas.
However, some differences are seen in the substructure of the jet-ambient gas which are seen better in 
the velocity map.

Figure \ref{fig:3Dvv} displays the 2D slices of the velocity field in Logarithm scale $v_p=\sqrt{v_{r}^2+v_{z}^2}$ 
in $r$-$z$ plane for run {\em HD3D} at different times.
It is seen that the  excited gas materials extend to the larger radii 
and have larger velocities compared to the reference run which affects the gas entropy as well.
Figure \ref{fig:ent3d} illustrates 2D slices of entropy of the gas in $r$-$z$ plane at different times.
It is observed that the entrained gas materials have larger entropy compared to the 2D reference run which
means that the larger fraction of energy is transfered in 3D run to the ambient gas.

The Probability density function of velocity can give a more accurate picture of the entrained gas evolution.
The PDF of velocity for 3D run is shown in Fig \ref{fig:pdf_3D}.
We recognize that the amount of (subsonic to supersonic) fluctuations increases compared to the reference run
and the excited gas materials gain higher velocities.
We observe that the peak of  the profile moves toward the larger velocities 
and the larger fraction of the gas materials stays in supersonic regime even at late evolutionary stage.

To investigate the accurate reasons of the differences seen in 3D run compared to axisymmetric run
the detailed studies are required which would be the aim of another paper.
However, we should consider some reasons which may play role.
 
It is known from the computational fluid dynamics that to solve the Navier-Stokes equations,
the act of discretizing the equations necessarily replaces terms in the equations with approximations. 
These approximations can be represented as an additional diffusive term, which is referred to in the literature
as “numerical diffusion,”  or “numerical viscosity.”
We should notice that in 3D run, we apply a lower resolution (to save computing time) which makes the equilibrium more unstable.
In particular, the lower resolution implying a somewhat higher numerical diffusivity (viscosity) that
results in the numerical heating and thus increases the entropy of the gas materials.
Consequently, the amount of the fluctuations in the surrounding gas increases.

Another reason may refer to the stability of the jet-ambient gas system in three dimensions.
There are some literature have studied the jet stability applying 3D hydrodynamic simulations \citep{2002ApJ...572..713H}
or 3D  MHD simulations \citep{2000ApJ...543..161X, 2013MNRAS.429.2482P,2013PPCF...55l4038K,2013MNRAS.436.1102M}.
It is known from instability studies, that some of the instabilities develop differently in three dimensions 
(like Magneto-rotational instability \citep{2016MNRAS.456.3782R}).
In addition, some substructures and non-axisymmetric effects are observable just in 3D simulations.
In our model setup, since the jet is in pressure equilibrium with the surrounding gas the jet 
interacts by shear motions with the surrounding gas.
By doing the 3D simulation of jet-ambient gas the shear motions in azimuthal direction strengthen 
which influence the stability of the system.  
In fact, the shear instability is not really considered in the axisymmetric run.

In total, we find that by using the full 3D simulation the larger fraction of 
the fluctuations are produced in the surrounding gas compared to the axisymmetric run
and the entrained gas gains higher velocities in a 3D run.
In particular, it is seen that the larger fraction of the gas materials stays 
in supersonic regime even at late evolutionary stage in the 3D run.

\begin{figure*}
\centering
\includegraphics[width=18cm]{\figurepath/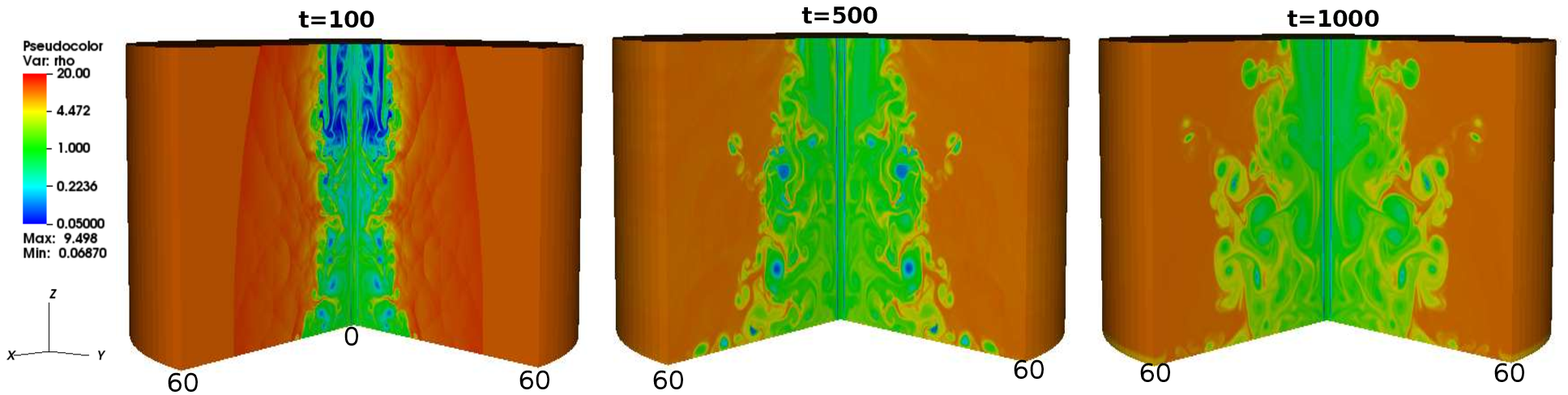}
  \caption{ 3D evolution of jet-ambient gas system. 
  Shown are the cuts of the mass density distribution in three dimensions
  for run {\em HD3D} at times $t= 100, 500, 1000$.}
\label{fig:run3D_rho}
\includegraphics[width=18cm]{\figurepath/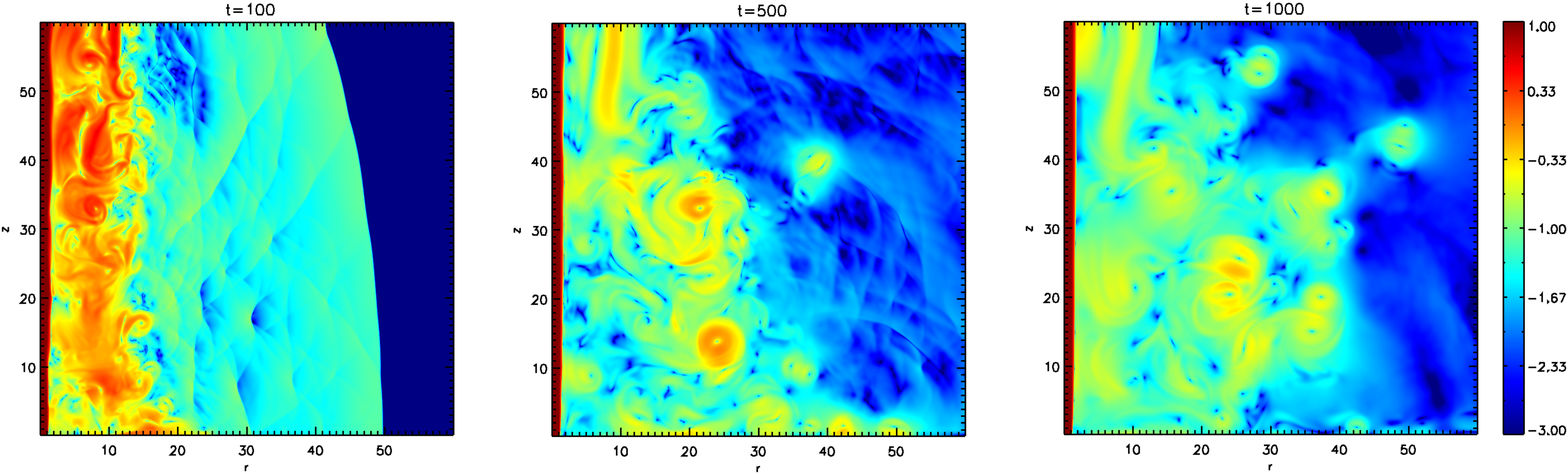}
 \caption{\sn Shown are  2D slices of the velocity field in Logarithm scale $v_p=\sqrt{v_{r}^2+v_{z}^2}$ 
in $r-z$ plane for run {\em HD3D} at times $t= 100, 500, 1000$.}
\label{fig:3Dvv}
\includegraphics[width=18cm]{\figurepath/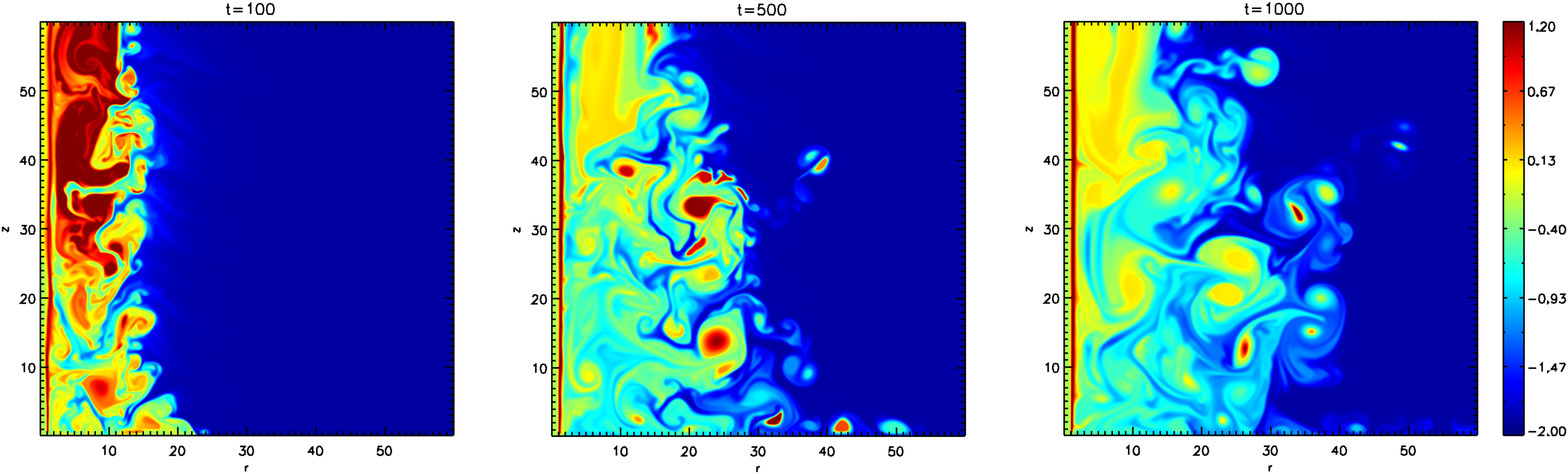}
  \caption{\sn Shown are 2D slices of the entropy in Logarithm scale in $r-z$ plane for run {\em HD3D} at  
  times $t= 100, 500, 1000$.}
\label{fig:ent3d}
\end{figure*}

\begin{figure}
\centering
\includegraphics[width=1.\columnwidth]{\figurepath/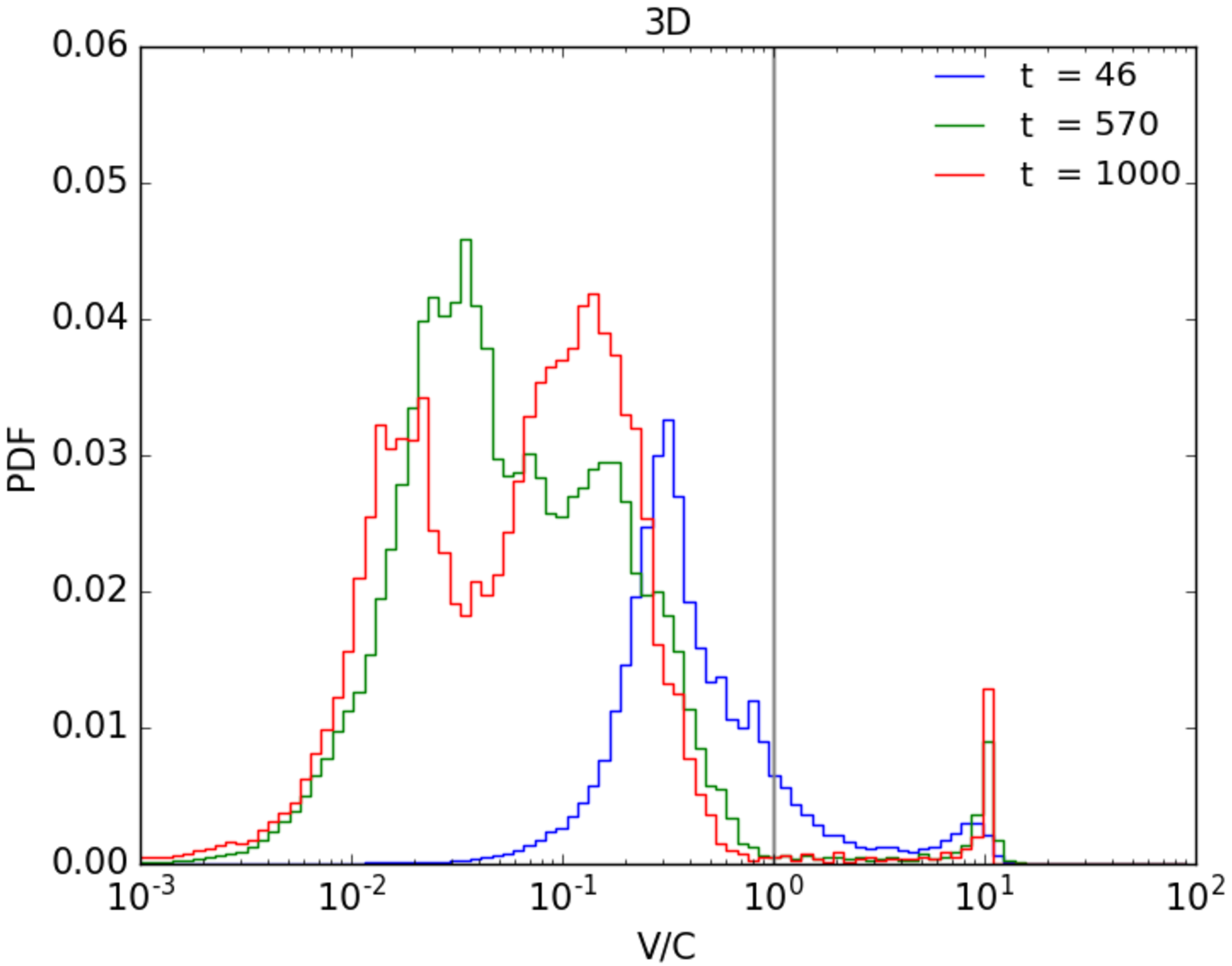}
 \caption{Shown are the plots of the probability Density Function of velocity for run {\em HD3D}
  at times 50, 500, 1000.}
\label{fig:pdf_3D}
\end{figure}

}

\section{Conclusion}
 In this work, we have presented a detailed study of feedback from protostellar jets based on 
simulations of a polytropic supersonic jet interacting with the denser
but in pressure equilibrium with the interstellar gas.
The presented simulations are performed in 2D and 3D applying the HD and MHD module of PLUTO 4.2 code.
The main question we address is whether the stellar jets can deliver the turbulence into the surrounding gas.
In addition, we investigate the most effective circumstances in which the driven turbulence is larger and can
survive for a long time scale.

We peresented a case study of different parameters runs including the jet Mach number, the initial jet velocity field 
and the background magnetic field geometries and the interacting jets.
In addition, we studeid the environmental effects on the jet-gas interaction by considering
the non-homogeneous surrounding gas containing the clumps in the model setup.

We used a statistical measurement of velocity, namely, the velocity probability density function (PDF), 
to quantify whether the main contributions of the excited motions, stays in subsonic or supersonic regime.

We have obtained the following results.\\

(1) We performed runs with different jet Mach numbers covering the range of transonic to highly supersonic stellar jets.
We found a clear correlation between the excited gas and the jet Mach number.
By increasing the jet Mach number a faster continues jet is injected into the outer environment and the
stronger interaction with ambient gas happens.
We recognized that a more powerful jet is a better candidate to drive and
maintain the turbulence in the surrounding gas.
This result is different from some previous works since we study the jet-ambient gas at different time scale and for a different 
length scale. \\

(2)In addition, we performed runs with different jet velocity distributions to investigate which velocity components 
has a larger impact on the jet-ambient gas interaction.
We found that the jet rotation produces the larger shearing motion which results in a more
excited gas in the interstellar gas.
As we mentioned in the model setup the jet is in pressure equilibrium with the surrounding gas initially.
Thus, the jet is stable itself and mostly interact by shear instabilities with the surrounding gas.
Including the jet rotation increases the shearing motion between the jet and the ambient gas and 
produces more velocity fluctuations or turbulence in the entrained gas.\\

(3) Moreover, we investigated the effects of the background  magnetic field on the turbulence driven by stellar jets.
Especially, we found that in run with the toroidal background magnetic field, the larger interaction of jet material and the ambient
gas happens and the more energy input is produced which is most probably due to the larger shearing motion between the jet and the ambient gas. \\

(4) Also, we studied the environmental effects on driving process of turbulence by stellar jets.
We performed simulations of jet interacting with the non-homogeneous environment containing the clump. 
We defined two different clumps, one with the higher density but in pressure equilibrium with its neighboring gas
called `` quiescent clump'' and  the one called ``explosive clump '' with over density and over pressure with respect 
to the neighboring materials.
In general, we found that the environmental effects are important in the evolution and maintenance of the turbulent motion
in the interstellar gas.
In particular, the run containing the long lasting clump ``quiescent clump'' provides the circumstances in which
the driven turbulence can be maintained for a longer time scale.\\

{(5) To study the efficiency of the stellar outflow as a turbulence driver on a larger scale (like cluster environment),
we performed simulations of interacting jets propagating in a larger computational domain. 
We found that the amount of subsonic and transonic fluctuations enhances but the supersonic 
turbulence is suppressed and can not be maintained for a long time scale after interaction between jets happens.\\

(6)In order to have a more realistic setup of jet-ambient gas interaction 
we extended our  model setup to three dimensions and performed a 3D simulation in HD regime.
We found that the 3D run displays the similar global evolution of jet to our 2D reference run but
some differences are seen in the substructure of the jet-ambient gas.
We found that the amount of (subsonic to supersonic) fluctuations increases
and the excited gas materials gain higher velocities in the 3D run.
Also, we recognized that the larger fraction of the gas materials stays in supersonic regime even at late evolutionary stage
compared  to the axisymmetric run.
}

Consequently, we confirm the previous studies on turbulence driven by protostellar jets that they induce the turbulence 
on neighboring region and are not the proper drivers of large-scale supersonic 
turbulence in molecular clouds. 
Although, they are considerable candidate in driving turbulent motions into the surrounding gas and in a smaller scale.

\acknowledgements
We thank Andrea Mignone and the PLUTO team for the possibility to use their code.
We thank Christian Fendt for valuable comments.
Our simulations were performed on the IPM cluster of the institute for research in fundamental sciences
and Isaac Cluster of Max Planck institute for Astronomy (MPIA).
This work was financed by the institute for research in fundamental sciences (IPM).
We thank an unknown referee for suggestions that improved the presentation of the paper.

\bibliographystyle{apj}
\bibliography{mybib}

\end{document}